\documentclass[11pt]{article}

\usepackage[utf8]{inputenc}
\usepackage[english]{babel}
\usepackage{xspace}
\usepackage{amsmath}
\usepackage{shuffle}
\usepackage{amssymb}
\usepackage{longtable}
\usepackage{multirow}
\usepackage{graphicx}
\usepackage{enumerate}
\usepackage{latexsym}
\usepackage{url}

\newcommand{\dfa}{DFA\xspace}
\newcommand{\nfa}{NFA\xspace}
\newcommand{\nfas}{NFAs\xspace}
\newcommand{\dfas}{DFAs\xspace}
\newcommand{\lcm}{\operatorname{lcm}}
\newcommand{\comp}[1]{\overline{#1}}

\newcommand{\shift}[1]{#1^{CS}}

\newcommand{\ignore}[1]{\relax}

\title{A Survey on Operational State Complexity}
\author{Yuan Gao$^{1}$ \and Nelma Moreira$^{2}$ \and Rog\'erio Reis$^{2}$ \and Sheng Yu$^{1}$\\
$^1$Department of Computer Science, University of Western Ontario\\
London, Ontario, Canada\\
$^2$CMUP \& DCC, Faculdade de Ciências da Universidade do Porto \\
  Rua do Campo Alegre, 4169-007 Porto, Portugal}

\begin{document}
\maketitle

\begin{abstract}
Descriptional complexity is
 the study of the conciseness of the various models representing formal languages. The state complexity of a regular language is the size, measured by the number of states of the smallest, either deterministic or nondeterministic, finite automaton that recognises it. Operational state complexity is the study of the state complexity of operations over languages.
In this survey, we review the state complexities of individual
regularity preserving language operations on regular and some
subregular languages. Then we revisit the state
complexities of the combination of individual operations. We also review methods of
estimation and approximation of state  complexity of more complex
combined operations.
\end{abstract}

\section{Introduction}
\label{sec:intro}
Automata theory is one of the oldest research areas in computer
science. Much research has been done on automata theory since 1950's. Work in many subareas of automata theory is still ongoing these days due to its new applications in areas such as software engineering, programming languages, parallel programming, network security, formal verification and natural language and speech processing \cite{Mo96,PeRi96,MoAlRoRoCoDe00,Sc06,LuYu09,wang13:_handb_of_finit_state_based}. 

Descriptional complexity and, in particular, state complexity is one of such active subareas. Generally speaking, the study of complexity mainly focuses on the following two kinds of issues: time and space complexity issues, i.e.\ time and space needed for the execution of the processes; or descriptional complexity issues, i.e.\ the succinctness of the model representations \cite{yu01:_state_compl_of_regul_languag}. In general, having succinct objects will improve our control on software, which may become smaller, more efficient and easier to certify.

State complexity is a type of descriptional complexity based on the finite machine model, and, in the domain of regular languages, it is related to the basic question of how to measure the size of a finite automaton. For the deterministic finite automaton (\dfa) case, the three usual answers are: the number of states, the number of transitions, or a combination of the two~\cite{yu01:_state_compl_of_regul_languag}. 
For a complete \dfa, whose transition function is defined for every state and every possible input symbol, the number of transitions is linear with the number of states, for each fixed alphabet.
Thus, the number of states becomes the key measure for the size of a complete \dfa. 
When considering the descriptional complexity of nondeterministic finite automata (\nfa), because this notion of completeness is not present, the measures based on the number of states and on the number transitions, are much more loosely related.

Since a regular language can be accepted by many \dfas with different number of states but only by one unique minimal, complete \dfa, the deterministic state complexity of a regular language is defined as the number of states of the minimal, complete \dfa accepting it. If we replace the minimal, complete \dfa with minimal \nfa, we have the definition of nondeterministic state complexity. Since state complexity is used as a natural abbreviation of deterministic state complexity by most researchers working in the area, we also follow the convention in this paper.

Complexity can be studied in two different flavours: in the worst case~\cite{yu01:_state_compl_of_regul_languag} and in the average case~\cite{nicaud99:_averag_state_compl_of_operat_unary_autom}. The worst-case complexity of a  class of regular languages is the supremum of the complexities of all the languages in the
class~\cite{yu01:_state_compl_of_regul_languag} whereas the average-case complexity, it is the average value of the complexities of those languages. Although its evident practical importance, there is still very few research on average-case state complexity. For that reason, in this paper, we mainly review worst-case results.

Results on descriptional complexity can be, roughly, divided into representational (or transformational) and operational. Representational complexity studies the complexity of transformations between models, by comparing the sizes of different representations of formal languages~\cite{Sa09-UWO-talk}. For example, given an $n$-state \nfa for a regular language, the \dfa which is equivalent to it has at most $2^n$ states, and this result, established in 1957, is considered the first state complexity result~\cite{rabin59:_finit_autom_and_their_decis_probl}. Operational state complexity studies the state complexity of operations on languages. When we speak about the state complexity of an operation on regular languages, we mean the state complexity of the class of resulting languages from the operation~\cite{yu01:_state_compl_of_regul_languag}. For example, when we say the state complexity of the intersection operation on two regular languages, accepted by $m$-state and $n$-state \dfas, respectively, is $mn$, we mean that $mn$ is the worst-case state complexity of the class of regular languages that can be represented as the intersection of an $m$-state \dfa language and an $n$-state \dfa language. Note that this implies that the intersection of any $m$-state \dfa language $L_1$ and $n$-state \dfa $L_2$ language has a \dfa with at most $mn$ states (upper bound) and that there exist languages $L_1$ and $L_2$ such that the minimal DFA for $L_1 \cap L_2$ has exactly $mn$ states (lower bound).

In this survey, we mostly concentrate in operational state complexity results. 
Although first studies go back to the 1960's and 1970's, research in
the area has been most active in the last two decades.  This can be
partially explained by the fact that back then, descriptional
complexity issues were not a priority for applications, as they are
today. But, also, due to its combinatorial nature many of the current
research is only possible with the help  of new high-performance symbolic manipulation software and powerful computers~\cite{GaYu12}.

The paper is organized as follows. After some preliminares in the next section, the notions of deterministic and nondeterministic state complexity  are considered in Section~\ref{sec:scnsc}. To better understand the possible gap between both measures is a main topic of research.
In Section~\ref{sec:sciop}, we review the state complexities of individual regularity preserving language operations, like,  Boolean operations, catenation, star, reversal, shuffle, orthogonal catenation, proportional removal, and cyclic shift, etc. These individual operations are fundamental and important in formal languages and automata theory research and applications. Results in these two sections are given for different classes of (sub)regular languages, e.g. general infinite, finite, unary, star-free, etc. In Section~\ref{sec:scco},  we revisit the state complexities of combined operations which are combinations of individual operations, e.g., star of union, star of intersection, star of catenation, star of reversal, union of star, intersection of star, etc. The state complexities of most of these combined operations are much lower than the mathematical composition of the state complexities of their component individual operations. We also review the methods of estimation and approximation of state complexity of combined operations which can be used for very complex combined operations.
Section~\ref{sec:conclusions} concludes this survey with some discussion on the results presented, highlighting some open problems and directions of future research.

\section{Preliminaries}
\label{sec:preliminares}
Here we recall some basic definitions related to finite
automata and regular languages. For a more complete presentation
the reader is referred to~\cite{yu97:_handb_formal_languag}.

The set of natural numbers is denoted by $\mathbb{N}$ and for $i, j \in \mathbb{N}$,
$[i, j] = \{ x \in \mathbb{N} \mid i \leq x \leq j \}$. The power set of a set $S$ is denoted by $2^S$ and the cardinality of
a finite set $S$ is $|S|$. 
In the following, $\Sigma$ stands always for a finite alphabet,
the empty word is represented by $\varepsilon$ and the set of all words
over $\Sigma$ by $\Sigma^\star$. A language is a subset
of $\Sigma^\star$. We say that $L \subseteq \Sigma^\star$ is a unary
(respectively, binary, ternary) language if
$|\Sigma| = 1$ (respectively, $|\Sigma| = 2$, $|\Sigma| = 3$).
Note this definition does not require that all symbols of
$\Sigma$ actually appear in words of $L$ and hence every unary
language is also a binary language and a binary language is always
a ternary language. A language $L$ is said to be finite if  $L$
is a finite subset of $\Sigma^\star$.

A {\em nondeterministic finite automaton\/} (\nfa) is a tuple
$A = (Q, \Sigma, \delta, q_0, F)$ where $Q$ is a finite set of states,
$\Sigma$ is a finite alphabet, $\delta : Q \times \Sigma \rightarrow 2^Q$
is the (multi-valued) transition function, $q_0 \in Q$ is the initial
state and $F \subseteq Q$ is the set of final (accepting) states.
The transition function is extended as a function
$\widehat{\delta} : Q \times \Sigma^\star \rightarrow Q$ by setting
$\widehat{\delta}(q,\varepsilon) = q$ for $q\in Q$ and for $w \in \Sigma^\star$,
$x \in \Sigma$, $\widehat{\delta}(q,wx) = \delta( \widehat{\delta}(q, w),x)$. To simplify notation, we  denote $\widehat{\delta}$ 
by $\delta$. The language recognized by the \nfa $A$ is 
$
L(A) = \{ w \in \Sigma^\star \mid \delta(q_0, w) \cap F \neq \emptyset \}.
$

An \nfa $A = (Q, \Sigma, \delta, q_0, F)$ is a
{\em complete deterministic finite automaton\/} (\dfa) if the transition function
$\delta$ is one-valued, that is, $\delta$ is a function
$Q \times \Sigma  \rightarrow Q$. An {\em incomplete} \dfa allows
the possibility that some transitions may be undefined, that is,
$\delta$ is a partial function $Q \times \Sigma \rightarrow Q$.

Both the \dfas and the \nfas define the class of regular languages~\cite{yu97:_handb_formal_languag}.
It is well known that any regular language has a unique
minimal (complete or incomplete) \dfa, that is, a unique
\dfa with the smallest number of states. For a given regular
language the sizes of the minimal, complete \dfa and minimal, incomplete
\dfa differ by at most one state. Furthermore, for a given \dfa
there exists an $n \log n$ time algorithm to compute
the minimal \dfa~\cite{yu97:_handb_formal_languag}. On the other hand, for a given regular language  there may
be more than one minimal \nfa and \nfa minimization is
PSPACE-hard~\cite{holzer11:_descr_and_comput_compl_of,yu97:_handb_formal_languag}.


\newtheorem{theo}{Theorem}
\newtheorem{propos}{Proposition}

\section{State Complexity and Nondeterministic State Complexity}
\label{sec:scnsc}
The \emph{state complexity} of a regular language $L$, $sc(L)$, is the
number of states of its minimal \dfa.  The \emph{nondeterministic state
complexity} of a regular language $L$, $nsc(L)$, is the number of
states of a minimal \nfa that accepts $L$. Since a \dfa is in particular an \nfa, for any regular language $L$
one has $sc(L)\leq nsc(L)$. It is well known that any $m$-state \nfa
can be converted, via the \emph{subset construction}, into an equivalent
\dfa with at most
$2^m$ states~\cite{rabin59:_finit_autom_and_their_decis_probl}
(we call this conversion \emph{determination}). Thus, $sc(L)\leq 2^{nsc(L)}$.
\begin{figure}[htb]
\centering
  \begin{tabular}[h]{lc}
{\small(i)}&\raisebox{-.7\height}{\includegraphics[width=9.5cm]{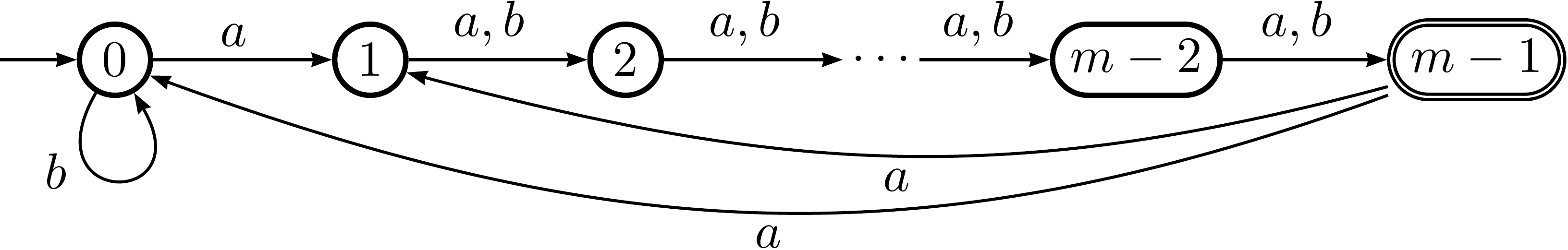}}
\ignore{\SmallPicture{\VCDraw{
      \begin{VCPicture}{(-2,-3)(17,1)}
       \State[0]{(0,0)}{0}\Initial{0}
       \State[1]{(3,0)}{1}
       \State[2]{(6,0)}{2}
       \SetStateLineStyle{none}
       \State[\cdots]{(9,0)}{3}
       \SetStateLineStyle{solid}
       \StateVar[m-2]{(12,0)}{m2}
       \FinalStateVar[m-1]{(16,0)}{m1}
       \EdgeL{0}{1}{a}\LoopS{0}{b}
       \EdgeL{1}{2}{a,b}
       \EdgeL{2}{3}{a,b}
       \EdgeL{3}{m2}{a,b}
       \EdgeL{m2}{m1}{a,b}
       \ArcL[.5]{m1}{1}{a}
       \VArcL[.5]{arcangle=20}{m1}{0}{a}
     \end{VCPicture}
   }
 }}\\
{\small(ii)}&
\raisebox{-.5\height}{\includegraphics[width=8.5cm]{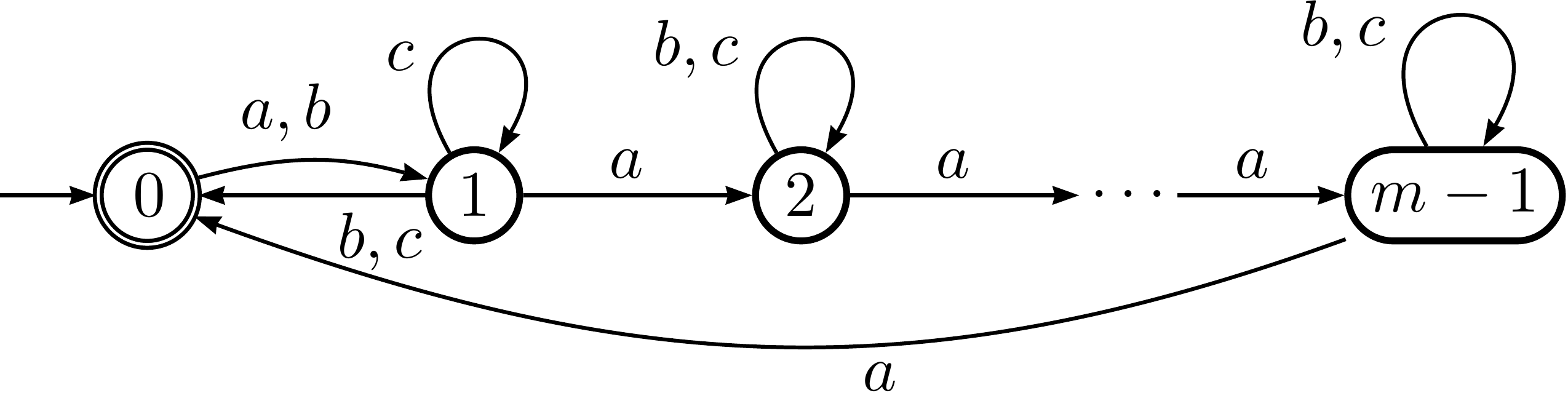}}
\ignore{\SmallPicture{\VCDraw{
    \begin{VCPicture}{(-2,-2)(15,2)}
      \FinalState[0]{(0,0)}{0}\Initial{0}
      \State[1]{(3,0)}{1}
      \State[2]{(6,0)}{2}
      \SetStateLineStyle{none}
      \State[\cdots]{(9,0)}{3}
      \SetStateLineStyle{solid}
      \StateVar[m-1]{(12,0)}{m1}
      \ArcL{0}{1}{a,b}
      \EdgeL[.2]{1}{0}{b,c}
      \LoopN{1}{c}
      \EdgeL{1}{2}{a}
      \LoopN{2}{b,c}
      \EdgeL{2}{3}{a}
      \EdgeL{3}{m1}{a}
      \LoopN{m1}{b,c}
      \VArcL{arcangle=20}{m1}{0}{a}
    \end{VCPicture}
  }
}}
\\{\small(iii)}&
\raisebox{-.5\height}{\includegraphics[width=10cm]{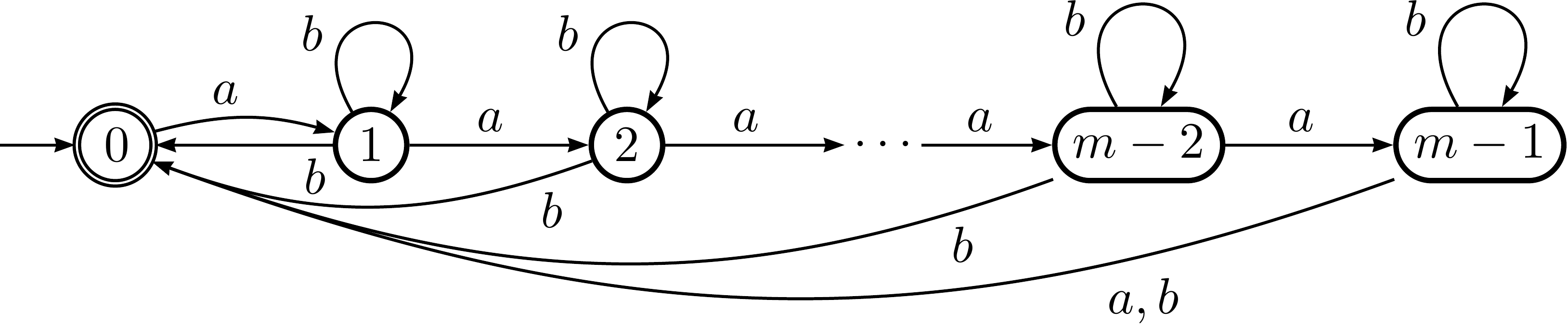}}
\ignore{\SmallPicture{\VCDraw{
    \begin{VCPicture}{(-2,-3)(18,2)}
      \FinalState[0]{(0,0)}{0}\Initial{0}
      \State[1]{(3,0)}{1}
      \State[2]{(6,0)}{2}
      \SetStateLineStyle{none}
      \State[\cdots]{(9,0)}{3}
      \SetStateLineStyle{solid}
      \StateVar[m-2]{(12,0)}{m2}
      \StateVar[m-1]{(16,0)}{m1}
      \ArcL{0}{1}{a}
      \EdgeL[.1]{1}{0}{b}\LoopN{1}{b}\EdgeL{1}{2}{a}
      \EdgeL{2}{3}{a}\VArcL[.1]{arcangle=20}{2}{0}{b}\LoopN{2}{b}
      \EdgeL{3}{m2}{a}
      \LoopN{m2}{b}\VArcL[.1]{arcangle=20}{m2}{0}{b}\EdgeL{m2}{m1}{a}
      \LoopN{m1}{b}\VArcL[.2]{arcangle=20}{m1}{0}{a,b}
    \end{VCPicture}
  }
 }}

\end{tabular}    
  \caption{\small{Moore {\small(i)}, Lupanov {\small(ii)}, and Meyer \&
    Fischer {\small(iii)} minimal $m$-state \nfas
    with equivalent minimal $2^m$-state \dfas}}
 \label{fig:nfadfaMooreLupanovMeyerFisher}
\end{figure}
To show that this upper bound is tight one must exhibit a family of languages
$(L_m)_{m\geq 1}$ such that $nsc(L_m)=m$ and $sc(L_m)=2^m$, for every
$m\geq 1$.
In 1963, Lupanov~\cite{lupanovn):_compar_of_two_types_of_finit_sourc}
showed that this upper bound is tight using a family of ternary
languages.  In 1971, Moore~\cite{moore71:_bound_for_state_set_size} and
Meyer and Fischer~\cite{meyer71:_econom_of_descr_by_autom} presented
different families of binary languages. All three families of \nfas are
represented in Figure~\ref{fig:nfadfaMooreLupanovMeyerFisher}.
However, for unary languages that upper bound is not
achievable~\cite{lyubich64:_estim_for_optim_deter_of,chrobak86:_finit_autom_and_unary_languag,chrobak03:_errat_to}.
Chrobak~\cite{chrobak86:_finit_autom_and_unary_languag,chrobak03:_errat_to}
proved that if $L$ is a unary language with $nsc(L)=m$, then
$sc(L)=O(F(m))$ where
\begin{equation}
  \label{eq:fm}
F(m)=\max\{\lcm(x_1,\ldots,x_l)\mid x_1,\ldots, x_l\geq 1 \text{ and }
x_1+\cdots+x_l=m\}  
\end{equation}
\noindent is the Landau's function and $\lcm$ denotes the least common
multiple. It
is known that $F(m)=e^{\Theta{(\sqrt{m\ln m})}}$, so
$sc(L)=e^{\Theta{(\sqrt{m\ln m})}}$. This asymptotic bound is tight,
i.e., for every $m$ there exists a unary language $L_m$ such that
$nsc(L_m)\leq m$ and $sc(L_m)= F(m-1)$.  Other related bounds were
studied by Meregethi and
Pighizzini~\cite{mereghetti00:_optim_simul_between_unary_autom}. 

For a general finite language $L$,  if $nsc(L)=m$ then
$sc(L)=\Theta(k^{\frac{m}{1+\log k}})$, $k=|\Sigma| > 1$, and this bound is
tight~\cite{salomaa97:_nfa_to_dfa_trans_for}. In the case of finite
binary languages, $\Theta(2^{\frac{m}{2}})$ is a tight
bound. In 1973, Mandl~\cite{mandl73:_precis_bound_assoc_with_subset} had
already proved that, for any finite binary language $L$, if $nsc(L)=m$
then $sc(L)\leq 2 \cdot 2^{m/2}- 1$ if $m$ is even, and $sc(L)\leq
3\cdot 2^{\lfloor m/2\rfloor}-1$ if $m$ is odd, and that these bounds are
tight. Finally, for finite unary  languages, nondeterminism does not lead to
significant improvements. If $L$ is a finite unary language with
$nsc(L)=m$, then $sc(L)\leq
m+1$~\cite{mandl73:_precis_bound_assoc_with_subset,salomaa97:_nfa_to_dfa_trans_for}.

In Section~\ref{sec:subregularlanguages} the state complexity of determination of other subregular languages is reviewed. As it will be evident from the results in the following sections, the complexity of determination plays a fundamental role in the operational complexity and thus the importance of its study \textit{per se}. 

The possible gap between state complexity and nondeterministic state
complexity for general regular languages lead to the notion of
\emph{magic number} introduced in 2000 by Iwama \emph{et
  al.}~\cite{iwama00:_tight_bound_number_of_states,iwama00:_famil_of_nfas_which_need}.
A number $\alpha$, such that $\alpha\in [m,2^m]$, is \emph{magic} for
$m$ with respect to a given alphabet of size $k$, if there is no minimal
$m$-state \nfa whose equivalent minimal \dfa has $\alpha$ states. This
notion has been extensively researched in the last decade and has been
extended to other gaps between two state complexity
values~\cite{lyubich64:_estim_for_optim_deter_of,chrobak86:_finit_autom_and_unary_languag,jiraskova01:_note_minim_finit_autom,
geffert05:_non_deter_and_size_of,geffert07:_magic_number_in_state_hierar,
geffert07:_state_hierar_for_one_way_finit_autom,jirasek08:_deter_blow_ups_of_minim,jiraskova09:_magic_number_and_ternar_alpha,jiraskova11:_magic_number_and_ternar_alphab,holzer12:_magic_number_probl_for_subreg_languag_famil}.
We summarize here some of the obtained results.  The general observation is
that, apart from unary languages, magic numbers are hard to find.  For
binary languages, it was shown that if $\alpha=2^m-2^n$ or
$\alpha=2^m-2^n-1$, for $n\in
[0,m/2-2]$~\cite{iwama00:_tight_bound_number_of_states}, and
$\alpha=2^m-n$ for $n\in [5, 2m-2]$ and some coprimality condition
holds for $n$~\cite{iwama00:_famil_of_nfas_which_need}, then $\alpha$
is not magic. Also, for a binary alphabet, all numbers $\alpha \in
[m,m+2^{\lfloor m/3\rfloor}]$ have been shown to be
non-magic~\cite{jiraskova08:_state_compl_of_compl_stars}, which
improves previous results,
$[m,m^2/2]$~\cite{jiraskova01:_note_minim_finit_autom} and
$[m,2^{\sqrt[3]{m}}]$~\cite{geffert05:_non_deter_and_size_of}.
For ternary or quaternary regular languages, and for languages over an
alphabet of exponential growing size there are no magic
numbers~\cite{jiraskova01:_note_minim_finit_autom,jirasek08:_deter_blow_ups_of_minim,jiraskova09:_magic_number_and_ternar_alpha,jiraskova11:_magic_number_and_ternar_alphab}.
For the unary case, however, trivially all numbers between
$e^{(1+o(1))\sqrt{m\ln m}}$ and $2^m$ are
magic~\cite{lyubich64:_estim_for_optim_deter_of,chrobak86:_finit_autom_and_unary_languag,geffert07:_magic_number_in_state_hierar}.
Moreover, it has been shown that there are much more magic than
non-magic numbers in the range from $m$ to $e^{(1+o(1))\sqrt{m\ln
    m}}$~\cite{geffert07:_magic_number_in_state_hierar}.  In the case
of finite languages, partial results were obtained by Holzer
\emph{et
  al.}~\cite{holzer12:_magic_number_probl_for_subreg_languag_famil}.
All numbers $\alpha\in[m+1, (\frac{m}{2})^2 +\frac{m}{2} + 1]$,
 if $m$ even, and $\alpha\in[m+1, (\frac{m-1}{2})^2  + m + 1]$, if $m$
 is odd, are non-magic. Moreover, all numbers of the form
 $3\cdot2^{\frac{m}{2}-1} + 2^i -1$, with $m$ even, and
 $2^{\frac{m+1}{2}} + 2^i - 1$, with
 $m$ odd, for some integer $i\in [1,\lceil\frac{m-1}{2}\rceil]$
are non-magic. In the same paper, the magic number problem is also
studied for other subregular language classes.

\subsection{State Complexity versus Quotient Complexity}
\label{sec:scqc}

Quotient complexity, introduced in 2009 by
Brzozowski~\cite{brzozowski09:_quotien_compl_of_regul_languag,brzozowskiar:_quotien_compl_of_regul_languag},
coincides, for regular languages, with the notion of state complexity
but it is defined in terms of languages and their (left) quotients.  The
\emph{left quotient} of a language $L$ by a word $w$ is defined as the
language $w^{-1}L = \{x \in \Sigma^\star \mid wx \in L\}$. The
\emph{quotient complexity} of $L$, denoted by $\kappa(L)$, is the number of distinct languages
that are left quotients of $L$ by some word. It is well known that, for a regular language $L$, the
number of left quotients is finite and is exactly the number of
states of the minimal \dfa accepting $L$. So, in the case of regular
languages, state complexity and quotient complexity
coincide. Considering that quotient complexity is given in terms of
languages, and their left quotients, some language algebraic
properties can be used in order to obtain upper bounds for the
complexity of operations over languages. Actually, the proof that the
set of
(left) quotients of a regular language is
finite~\cite{brzozowski64:_deriv_of_regul_expres} was one of the
earliest studies of state complexity.  Quotient complexity can also be
useful to show that an upper bound is tight.  If a given operation
can be expressed as a function of other operations (for example,
$L_1-L_2 = L_1\cap \comp{L_2}$), then, witnesses for the worst-case complexity of those
operations can be used to provide a witness for the complexity of the
first operation.

\section{State Complexity of Individual Operations}
\label{sec:sciop}

The \emph{state complexity of an operation} (or \emph{operational
  state complexity}) on regular languages is the worst-case state
complexity of a language resulting from the operation, considered as a
function of the state complexities of the operands. Adapting a formulation from Holzer and
Kutrib~\cite{holzer09:_nondet_finit_autom_recen_resul}, given a binary
operation $\circ$, the \emph{$\circ$-language operation state
  complexity problem} can be stated as follows:
 \begin{itemize}
 \item Given an $m$-state \dfa $A_1$ and an $n$-state \dfa  $A_2$.
 \item How many states are sufficient and necessary, in the worst case,
   to accept the language $L(A_1) \circ L(A_2)$ by a \dfa?
 \end{itemize}

 This formulation can be generalized for operations with other arities, other kinds of 
 automata and classes of languages.
  An upper bound can be obtained by providing an algorithm that, given
 \dfas for the operands, constructs  a \dfa that accepts the resulting
 language. The number of states of the resulting \dfa is an upper bound for the 
 state complexity of the referred operation.
 To show that an upper
 bound is tight, for each operand a family of languages (one language,
 for each possible value of the state complexity) must be given such
 that the resulting automata achieve that bound.  We can call those
 families \emph{witnesses}.  The same approach is used to obtain the
 nondeterministic state complexity of an operation on regular
 languages.  No proofs are here presented for the stated results,
 although several examples of families of languages, for which the
 operations achieve a certain upper bound, are given.
There are very few results of the study of state complexity on the
average case. However, whenever some results are known they are
mentioned together with the corresponding worst-case analysis.

 In this section, the following notation is used.  When considering
unary operations, let $L$ be regular language with $sc(L)=m$
($nsc(L)=m$) and let $A=(Q,\Sigma,\delta,q_0,F)$ be the complete
minimal \dfa (a minimal \nfa) such that $L=L(A)$. Furthermore,
$|\Sigma|=k$ or $|\Sigma|=f(m)$ if a growing alphabet is taken into
account, $|F|=f$, and $|F-\{q_0\}|=l$.  In the same way, for binary
operations let $L_1$ and $L_2$ be regular languages over the same
alphabet with $sc(L)=m$ ($nsc(L)=m$) and $sc(L_2)=n$ ($nsc(L_2)=n$),
and let $A_i=(Q_i,\Sigma,\delta_i,q_i,F_i)$ be complete minimal \dfas
(minimal \nfas) such that $L_i=L(A_i)$, for $i\in[1,2]$. Furthermore,
$|\Sigma|=k$ or $|\Sigma|=f(m,n)$ if a growing alphabet is taken into
account, $|F_i|=f_i$, and $|F_i-\{q_i\}|=l_i$, for $i\in [1,2]$.

\subsection{Basic Operations}

In this section we review the main results related with state
complexity (and nondeterministic state complexity) of some basic
operations on regular languages: Boolean operations (mainly union,
intersection, and complement), catenation, star (and plus), and
reversal. For some  classes of languages, left and right quotients 
are also
considered. Because their particular characteristics, that were
already pointed out in Section~\ref{sec:scnsc}, for each operation the
languages are divided into \emph{regular} ($k\geq 2$
and infinite), \emph{finite} ($k\geq 2$), \emph{unary } (infinite) and
\emph{finite unary}. Some other subregular languages are considered in
Section~\ref{sec:subregularlanguages}. Whenever known, results on the range of
complexities that can be reached for each operation are also
presented. This extension of the notion of magic number to operational state complexity is now an active topic of research.

There are some other survey papers that partially review the results here
presented and that were a reference to our
presentation~\cite{yu97:_handb_formal_languag,yu01:_state_compl_of_regul_languag,yu02:_state_compl_of_finit_and,hromkovic02:_descr_compl_of_finit_autom,yu05:_state_compl,salomaa07:_descr_compl_of_nondet_finit_autom,holzer09:_nondet_finit_autom_recen_resul,brzozowski09:_quotien_compl_of_regul_languag,holzer09:_descr_and_comput_compl_of_finit_autom,holzer11:_descr_and_comput_compl_of}.

\subsubsection{General Regular Languages}
\label{sec:generalregularbop}
Table~\ref{tab:scnscregular} summarizes the results for general
regular languages. The (fifth) third column contains the smallest alphabet
size of the witness languages for the (nondeterministic) state
complexity given in the (fourth) second column, respectively.  Columns
with this kind of information also appear in several tables to
follow.
 
\begin{table}[htbp]
\begin{tabular}{|l||c|c|c|c|}\hline
\multicolumn{5}{|c|}{Regular}\\\hline
&\multicolumn{1}{c}{sc}&\multicolumn{1}{c|}{$|\Sigma|$}&\multicolumn{1}{c}{nsc}&\multicolumn{1}{c|}{$|\Sigma|$}\\\hline\hline
$L_1\cup L_2$&$mn$&2&$m+n+1$&2\\\hline
$L_1\cap L_2$&$mn$&2&$mn$&2\\\hline
\mbox{$\comp{L}$}&$m$&1&$2^{m}$&2\\\hline
  $(L_1- L_2)$&$mn$ &2&&\\\hline
 $(L_1\oplus L_2)$&$mn$&2&&\\\hline
\multirow{2}{*}{$L_1L_2$}&$m2^n-f_12^{n-1}$, if $\ n>1$&2&\multirow{2}{*}{$m+n$}&\multirow{2}{*}{2}\\\cline{2-3}
&$m$, if $n=1$& 1&&\\\hline
\multirow{3}{*}{$L^\star$}&
$2^{m-1}+2^{m-l-1}$, if $m>1,\, l>0$&2&\multirow{3}{*}{$m+1$}&\multirow{3}{*}{2}\\\cline{2-3}
&$m$, if $m>1,\, l=0$&1&&\\\cline{2-3}

&$m+1$, if $m=1$&1&&\\\hline

$L^{+}$&$2^{m-1}+2^{m-l-1}-1$&2&$m$&2\\\hline
$L^{R}$&$2^{m}$&2&$m+1$&2\\\hline
$L_2\setminus L_1$&$2^m-1$&2&&\\\hline
$L_1\,/\,L_2$&$m$&1&&\\\hline
$w^{-1} L$&$m$&$1$&$O(m+1)$&\\\hline
$Lw^{-1}$&$m$&1&$m$&1\\\hline
\end{tabular}
  \centering
  \caption{\small{State complexity and nondeterministic state complexity for basic operations on regular languages}}\label{tab:scnscregular}
\end{table}

In 1994, Yu \emph{et al.}~\cite{yu94:_state_compl_of_some_basic}
studied the state complexity of catenation, star, reversal, union,
intersection, and  left and right quotients.  They also studied the
state complexity of some operations for unary languages.  More than two decades before, in
1970, Maslov~\cite{maslov70:_estim_of_number_of_states} had presented
some estimates for union, catenation, and star. Although Maslov
considered possible incomplete \dfas, and the paper has some
incorrections, the binary languages presented are tight witnesses for
the upper bounds for that three
operations~\cite{brzozowski09:_quotien_compl_of_regul_languag}. Rabin and
 Scott~\cite{rabin59:_finit_autom_and_their_decis_probl}
indicated the upper bound $mn$ for the intersection (that also applies
to union). Maslov and Yu \emph{et al.} gave similar witnesses of
tightness, both for union and intersection. The families of languages
given by Yu \emph{et al.} for intersection are $\{x\in\{a,b\} \mid
\#_a(x) = 0\pmod{m}\}$ and $\{x\in\{a,b\} \mid \#_b(x)= 0
\pmod{n}\}$. Their complements are witnesses for
union. Hricko~\emph{et al.}~\cite{hricko05:_union_and_inter_of_regul}
showed that for any integers $m\geq 2$, $n\geq 2$, and $\alpha\in
[1,mn]$, there exist binary languages $L_1$ and $L_2$ such that
$sc(L_1) = m$, $sc(L_2) = n$, and $sc(L_1\cup L_2) = \alpha$. Thus, there are no magic numbers for the union. The same
holds for intersection.

Complementation for \dfas is trivial (one has only to exchange the final
states) and thus, the state complexity of the complement is the same one of the 
original language, i.e., $sc(\comp{L})=sc(L)$. For other Boolean
operations (set difference, symmetric difference, exclusive
disjunction, etc.) the state complexity can be obtained by expressing
them as a function of union, intersection and
complement~\cite{brzozowski09:_quotien_compl_of_regul_languag}.

\begin{figure}[htb]
   \centering
   \begin{tabular}[t]{lc}
    {\small(i)}& \raisebox{-.5\height}{\includegraphics[width=10cm]{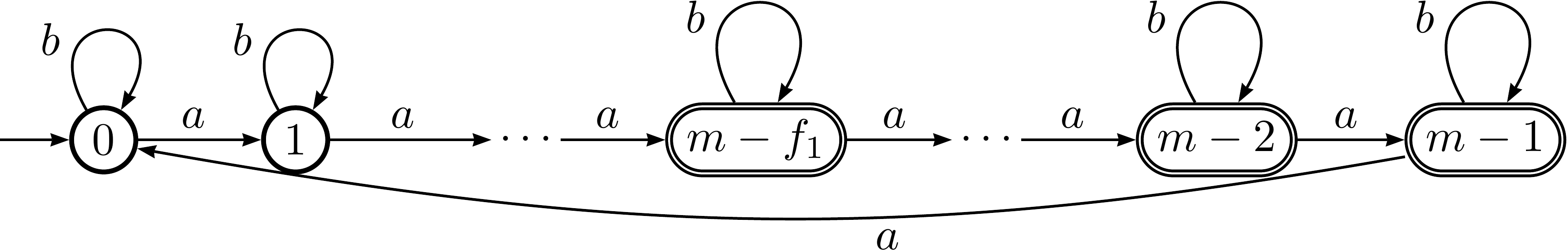}}
		\ignore{\SmallPicture{\VCDraw{
             \begin{VCPicture}{(-2,-2)(20,2)}
               \State[0]{(0,0)}{0}\Initial{0}
               \State[1]{(2.5,0)}{1}
               \SetStateLineStyle{none}
               \State[\cdots]{(5.5,0)}{2}
               \State[\cdots]{(11.5,0)}{mf1}
               \SetStateLineStyle{solid}
               \FinalStateVar[m-f_1]{(8.5,0)}{mf}
               \FinalStateVar[m-2]{(14.5,0)}{m2}
               \FinalStateVar[m-1]{(18,0)}{m1}
               \LoopN{0}{b}
               \LoopN{1}{b}
               \LoopN{mf}{b}
               \LoopN{m2}{b}
               \LoopN{m1}{b}
               \EdgeL{0}{1}{a}
               \EdgeL{1}{2}{a}
               \EdgeL{2}{mf}{a}
               \EdgeL{mf}{mf1}{a}
               \EdgeL{mf1}{m2}{a}
               \EdgeL{m2}{m1}{a}
               \VArcL{arcangle=10}{m1}{0}{a}
             \end{VCPicture}
           }
         }}
    \\{\small(ii)}& \raisebox{-.5\height}{\includegraphics[width=7.5cm]{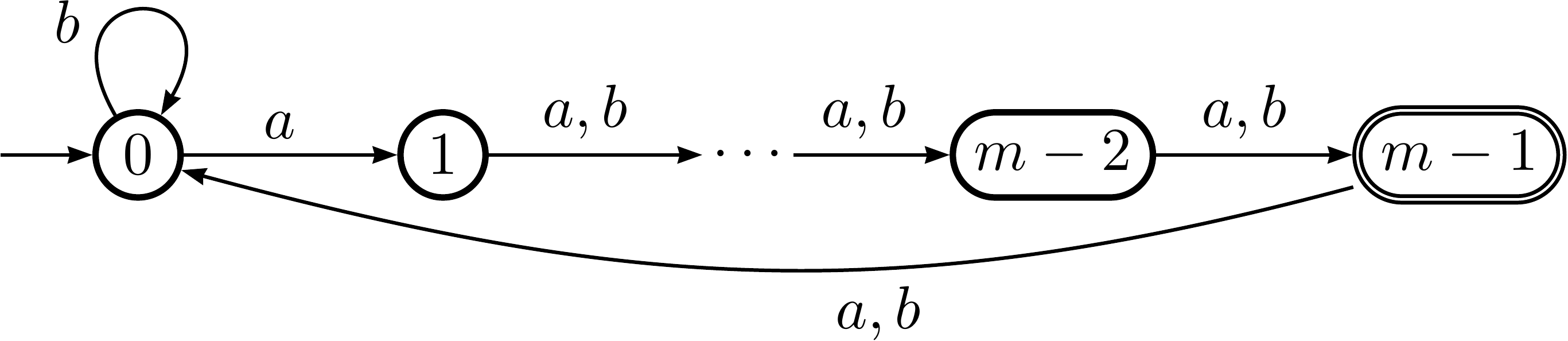}}
		\ignore{\SmallPicture{\VCDraw{
             \begin{VCPicture}{(-1,-2)(14,3)}
               \State[0]{(0,0)}{0}\Initial{0}
               \State[1]{(3,0)}{1}
               \SetStateLineStyle{none}
               \State[\cdots]{(6,0)}{2}
               \SetStateLineStyle{solid}
               \StateVar[m-2]{(9,0)}{m2}
               \FinalStateVar[m-1]{(13,0)}{m1}
               \LoopN{0}{b}\EdgeL{0}{1}{a}
               \EdgeL{1}{2}{a,b}
               \EdgeL{2}{m2}{a,b}
               \EdgeL{m2}{m1}{a,b}
               \ArcL{m1}{0}{a,b}
             \end{VCPicture}
           }
         }}

   \end{tabular}
   \caption{\small{Witness \dfas for all range of state complexities of the catenation}}
   \label{fig:witnesscatenationstar}
 \end{figure}

For catenation, Yu \emph{et al.} gave the upper bounds
$m2^n-f_12^{n-1}$, if $m\geq 1, n\geq 2$; and $m$, if $m\geq 1,
n=1$. They presented binary languages tight bound witnesses for $m\geq
1,\; n=1$ and $m=1,\; n\geq2$, but ternary languages tight bound
witnesses for $m> 1,\; n\geq 2$. However, the bound is tight for the
following binary language families presented by Maslov:
$\{w\in\{a,b\}^\star\mid \#_a(w)=(m-1)\pmod{m}\}$ and $L((a^\star
b)^{n-2}(a+b)(b+a(a+b))^\star)$, for all $m,n\geq 2$ and
$f_1=1$. Other families of binary languages for which the catenation
achieves the upper bound were presented by
Jir\'askov\'a~\cite{jiraskova05:_state_compl_of_some_operat}.  Concerning the possible existence of magic numbers, the
same
author~\cite{jiraskova09:_concat_of_regul_languag_and_descr_compl,jiraskova11:_concat_of_regul_languag_and_descr_compl}
proved that, for all $m$, $n$ and $\alpha$ such that either $n=1$ and
$\alpha\in [1,m]$, or $n\geq 2 $ and $\alpha\in[1,m2^n-2^{n-1}]$,
there exist languages $L_1$ and $L_2$ with $sc(L_1)=m$ and
$sc(L_2)=n$, defined over a growing alphabet, such that
$sc(L_1L_2)=\alpha$. Moreover, Jir\'asek \emph{et
  al.}~\cite{jirasek05:_state_compl_of_concat_and} showed that the
upper bound $m2^n - f_12^{n-1}$ on the catenation of two languages
$L_1$ and $L_2$, with $sc(L_1)=m\geq 2$ and $sc(L_2)=n\geq 2$
respectively, are tight for any integer $f_1$ with $f_1\in
[1,m-1]$. The witness language families are binary and accepted by the
\dfas presented in Figure~\ref{fig:witnesscatenationstar}.

\begin{figure}[htb]
  \centering

\includegraphics[width=8cm]{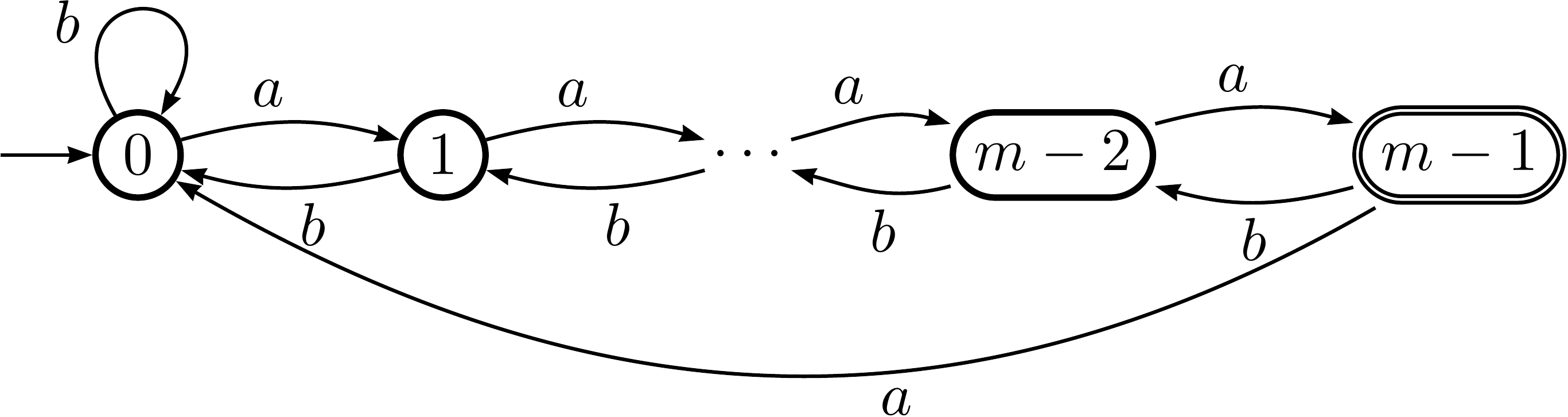}\ignore{\SmallPicture{\VCDraw{
      \begin{VCPicture}{(-1,-2.5)(14,3)}
        \State[0]{(0,0)}{0}\Initial{0}
        \State[1]{(3,0)}{1}
        \SetStateLineStyle{none}
        \State[\cdots]{(6,0)}{2}
        \SetStateLineStyle{solid}
        \StateVar[m-2]{(9,0)}{m2}
        \FinalStateVar[m-1]{(13,0)}{m1}
        \LoopN{0}{b}\ArcL{0}{1}{a}
        \ArcL{1}{2}{a}\ArcL{1}{0}{b}
        \ArcL{2}{1}{b}\ArcL{2}{m2}{a}
        \ArcL{m2}{m1}{a}\ArcL{m2}{2}{b}
        \VArcL{arcangle=30}{m1}{0}{a}\ArcL[.5]{m1}{m2}{b}
      \end{VCPicture}
     }
   }}

  \caption{\small{Maslov's witness \dfas for the state complexity of the
    star}}\label{fig:maslovstar}
\end{figure}

 The state complexity for the star on a regular language $L$ was
 studied by Yu \emph{et al.}. A lower bound of $2^{m-1}$ was presented
 before, by Ravikumar and
 Ibarra~\cite{ravikumar89:_relat_type_of_ambig_of,ravikumar90:_some_applic_of_techn_of}.
 If $sc(L)=1$ then either $L=\Sigma^\star$, and $sc(L^\star)=1$, or
 $L=\emptyset$, and $sc(L^\star)=2$. If $sc(L)=m>1$, but $l=0$,
 i.e., the minimal \dfa accepting $L$ has the initial state as the
 only final state, then $sc(L^\star)=m$, as $L=L^\star$. Finally, if
 $sc(L)=m>1$, and $l>0$, then $sc(L^\star)\leq
 2^{m-1}+2^{m-l-1}$. The upper bound $2^{m-1}+2^{m-2}$ is achieved
 for the language $\{w\in \{a,b\}^\star\mid \#_a(w) \text{ is odd}\}$, if
 $m=2$; if $m> 2$, for the family of binary languages accepted by
 the \dfas\label{starwitness} presented in
 Figure~\ref{fig:witnesscatenationstar}:(ii).
We note that although the upper bound given by Maslov is not correct
($\frac{3}{4}2^{m}-1$ instead of $\frac{3}{4}2^{m}$), the family of
languages he presented are witnesses for the above bound (for $m>2$). Those
languages are accepted by the  \dfas presented in Figure~\ref{fig:maslovstar}.

Jir\'askov\'a~\cite{jiraskova08:_state_compl_of_compl_stars} proved that for all integers $m$ and $\alpha$ with either $m=1$ and $\alpha \in
[1,2]$, or $m \geq 2$ and $\alpha \in [1, 2^{m-1} +2^{m-2}]$, there
exists a language $L$ over an alphabet of size $2^m$ such that
$sc(L) = m$ and $sc(L^\star) = \alpha$. This result was improved by~Jir\'askov\'a \emph{et al.}~\cite{jiraskova14:_kleen_closur_regul_and_prefix_free_languag} by using an alphabet of size atmost $2m$. Again, no gaps or magic numbers exist for the Kleene star operation. 

The state complexity for the plus on a regular language $L$ ($L^{+} = LL^\star$) coincides with the one for star in the first two cases,
but for $m>1$ and $l>0$ one state is saved (as a new initial state is
not needed).

\begin{figure}[htb]
  \centering

 \includegraphics[width=9cm]{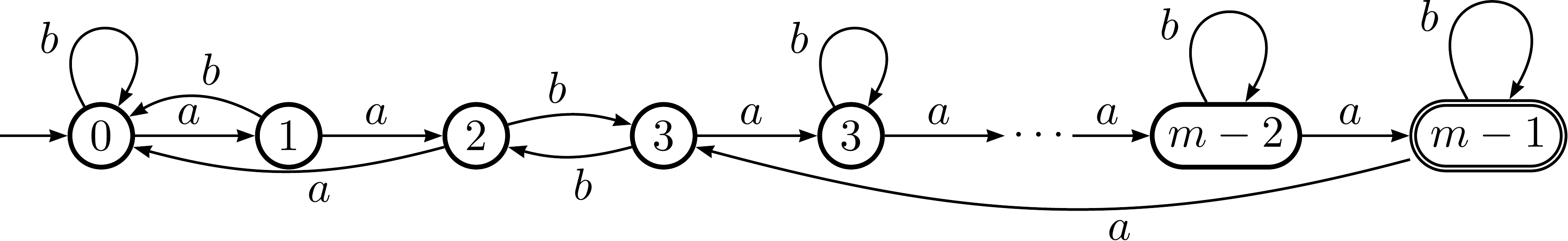}
 \ignore{\SmallPicture{\VCDraw{
        \begin{VCPicture}{(-2,-1)(18,2)}
          \State[0]{(0,0)}{0}\Initial{0}
          \State[1]{(2.5,0)}{1}
          \State[2]{(5,0)}{2}
          \State[3]{(7.5,0)}{3}
          \State[3]{(10,0)}{4}
          \SetStateLineStyle{none}
          \State[\cdots]{(12.5,0)}{5}
          \SetStateLineStyle{solid}
          \StateVar[m-2]{(15,0)}{m2}
          \FinalStateVar[m-1]{(18.5,0)}{m1}
          \LoopN{0}{b}\EdgeL{0}{1}{a}\ArcL{2}{0}{a} \VArcR{arcangle=-30}{1}{0}{b}
          \EdgeL{1}{2}{a}
          \ArcL{2}{3}{b}         \ArcL{3}{2}{b}
          \EdgeL{3}{4}{a}\ArcL{m1}{3}{a}
          \LoopN{4}{b}
          \EdgeL{4}{5}{a}
          \EdgeL{5}{m2}{a}         \LoopN{m2}{b}
          \EdgeL{m2}{m1}{a}
          \LoopN{m1}{b}
        \end{VCPicture}
      }
 }}

  \caption{\small{Witness \dfas for the state complexity of the reversal}}
  \label{fig:binarywitnessreversal}
\end{figure}

In 1966, Mirkin~\cite{mirkin66:_dual_autom} pointed out that the
reversal of the \nfas given by Lupanov as an example of a tight bound
for determination (see
Figure~\ref{fig:nfadfaMooreLupanovMeyerFisher}:(ii)), were
deterministic. This yields that $2^m$ is a tight upper bound for the
state complexity of reversal of a (at least ternary) language $L$
such that
$sc(L)=m$. Leiss~\cite{leiss85:_succin_repres_of_regul_languag}
studied also this problem and proved the tightness of the bound for
another family of ternary languages. Yu \emph{et al.} presented also
(independently) Lupanov example. Salomaa \emph{et
  al.}~\cite{salomaa04:_state_compl_of_rever_of_regul_languag} studied
several classes of languages where the upper bound is
achieved. Nevertheless, a family of binary languages therein presented
as meeting the upper bound for $m\geq 5$ was later shown not to be
so~\cite{jiraskova10:_compl_in_prefix_free_regul_languag}. A family of
binary languages for which the upper bound for reversal is tight was
given by Jir\'askov\'a and S\v ebej~\cite{regular10:_juraj_s_ebej,jiraskova11:_note_rever_of_binar_regul_languag} and their minimal
\dfas are represented in Figure~\ref{fig:binarywitnessreversal}.

In the paper cited
above~\cite{jiraskova08:_state_compl_of_compl_stars}, Jir\'askov\'a shown that for all $m$ and $\alpha$ with $2\leq m\leq \alpha\leq 2m$, there exists a binary languague $L$ such that $sc(L)=m$ and $sc(L^R)=\alpha$. Allowing alphabets of size $2^m$ and $m\geq 3$, the reversal operation has no magic numbers in the range~$[\log m, 2^m]$. This result was improved by S\v ebej~\cite{sebej13:_rever_regul_languag_and_descr_compl} considering an alphabet of size $2m-2$.  S\v ebej gives also some enhanced partial results for the binary case.


Yu \emph{et al.} showed that the state complexity for the left
quotient of a regular language $L_1$ by an arbitrary language $L_2$,
$L_2\setminus L_1$, is less or equal to $2^m-1$, with $sc(L_1)=m$, and
that this bound is tight for the family of binary languages given in
Figure~\ref{fig:witnesscatenationstar}:(ii)
and considering $L_2=\Sigma^\star$. In
1971, Conway~\cite{conway71:_regul_algeb_and_finit_machin}
had already stated that if $L_2$ is a regular language then
$sc(L_2\setminus L_1)\leq 2^m$. For the right quotient of a regular
language $L_1$ by an arbitrary language $L_2$ one has $sc(L_1/L_2)\leq
m$. The minimal \dfa accepting $L_1/L_2$ coincides with the one for
$L_1$, except that the set of final states is the set of states $q\in
Q_1$ such that there exists a word of $w\in L_2$ such that
$\delta_1(q,w)\in F_1$. The bound is tight for $L_2=\{\varepsilon\}$.
For the left and the right quotients of a regular language $L$ by a
word $w\in \Sigma^\star$ it is then easy to see that
$sc(w^{-1}L)=sc(Lw^{-1})\leq m$. As a family of languages for which
the upper bound is tight consider
$(a^m)^\star$
and
$w\in
\{a\}^\star$~\cite{ellul03:_descr_compl_measur_of_regul_languag}.

The state complexity of basic operations on \nfas was first studied
by Holzer and Kutrib~\cite{holzer03:_state_compl_of_basic_operat}, and
also by Ellul~\cite{ellul03:_descr_compl_measur_of_regul_languag}.  We
note that for state complexity purposes it is tantamount to consider
\nfas with or without $\varepsilon$-transitions. \nfas are considered
with only one initial state and trimmed, i.e., all states are
accessible from the initial state and from all states a final state is
reached.

For union, only a new initial state with $\varepsilon$ transitions for
each of the operands initial states is needed, thus $sc(L_1\cup
L_2)\leq m+n+1$. To see that the upper bound is tight, consider the
families $(a^m)^\star$ and $(b^n)^\star$ over a binary
alphabet. For intersection, a product construction is needed.

\begin{figure}[htb]
  \centering
\includegraphics[width=7cm]{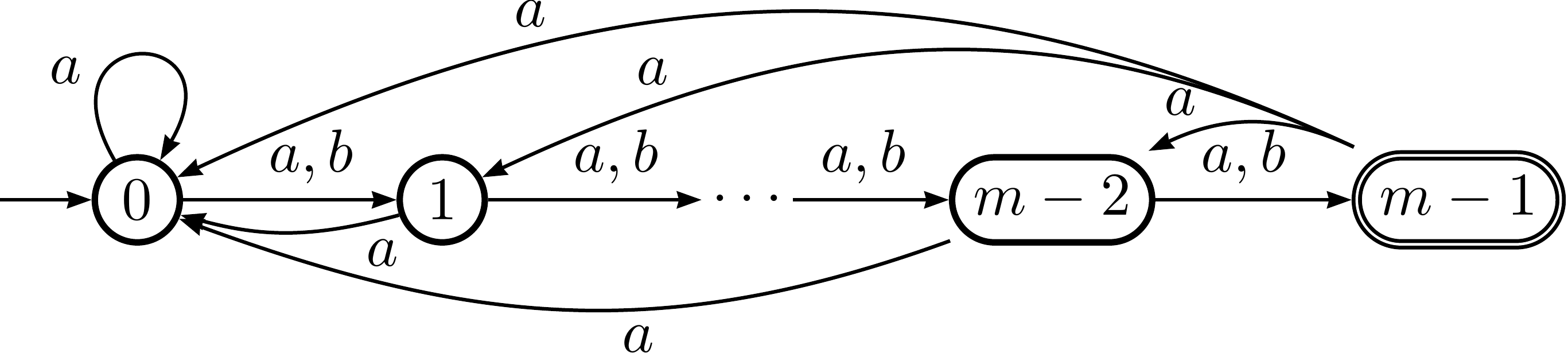}
\ignore{\SmallPicture{\VCDraw{
        \begin{VCPicture}{(-2,-2)(14,2)}
          \State[0]{(0,0)}{0}\Initial{0}
          \State[1]{(3,0)}{1}
          \SetStateLineStyle{none}
          \State[\cdots]{(6,0)}{2}
          \SetStateLineStyle{solid}
          \StateVar[m-2]{(9,0)}{m2}
          \FinalStateVar[m-1]{(13,0)}{m1}
          \LoopN{0}{a}
          \EdgeL[.6]{0}{1}{a,b}
          \EdgeL[.6]{1}{2}{a,b}
          \EdgeL{2}{m2}{a,b}
          \EdgeL{m2}{m1}{a,b}
          \ArcL[.1]{1}{0}{a}
          \VArcL{arcangle=20}{m2}{0}{a}
          \VArcR[.8]{arcangle=-25}{m1}{m2}{a}
          \VArcR[.8]{arcangle=-25}{m1}{1}{a}
          \VArcR[.7]{arcangle=-25}{m1}{0}{a}
        \end{VCPicture}
      }
    }}

  \caption{\small{Witness \nfas for the nondeterministic state
      complexity of complementation}}
  \label{fig:witnessnfas}
\end{figure}

The nondeterministic state complexity of the complementation is,
trivially, at most $2^m$. That this upper bound is tight even for
binary languages was proved by
Jir\'askov\'a~\cite{jiraskova05:_state_compl_of_some_operat}, using a
\emph{fooling-set lower-bound
  technique}~\cite{birget91:_inter_of_regul_languag_and_state_compl,glaister96:_lower_bound_techn_for_size,hromkovic97:_commun_compl_and_paral_comput}.
Those languages are accepted by the \nfas presented in
Figure~\ref{fig:witnessnfas} (for $m>2$).
%

See Holzer and Kutrib~\cite{holzer09:_nondet_finit_autom_recen_resul}
for other witness languages.  Using the same techniques, Jir\'askov\'a
and Szabari~\cite{jirasek05:_state_compl_of_concat_and} proved that  for all integers $m\geq 1$
  and $\alpha\in [\log m, 2^m]$, there exists a language $L$  over an alphabet of exponential
growing size, such
 that $nsc(L) = m$ and $nsc(\comp{L}) = \alpha$.  This result  was improved to a five-symbol alphabet by
Jir\'askov\'a~\cite{jiraskova08:_state_compl_of_compl_stars}. 


Mera and Pighizzini~\cite{mera05:_compl_unary_nondet_autom} proved a
related \emph{best case} result, i.e., for every $m\geq 2$ there
exists a language $L$ such that $nsc(L)=m$, $nsc(\comp{L})\leq
m+1$ and $sc(L)=sc(\comp{L})=2^m$. However, as we will see below,
this result does not hold if unary languages are considered.
 
The upper bound for the nondeterministic state complexity of
catenation is $m+n$ and this bound can be reached considering the
witness binary languages given for
union\label{witnessnsccatenation}. All the values $\alpha\in [1,m+n]$
can be obtained as nondeterministic state complexity of catenation of
unary
languages~\cite{jiraskova09:_concat_of_regul_languag_and_descr_compl}.

For the plus of a regular language $L$, we have $nsc(L^+)\leq
nsc(L)=m$: an \nfa accepting $L^+$ coincides with one accepting
$L$ except that each final state has also the transitions to the
initial state. In the case of the star, one more state can be needed
(if $L$ does not accept the empty word), i.e., $sc(L^\star)\leq
m+1$.  Witness languages of the tightness of these bounds are $\{w\in
\{a,b\}^\star\mid \#_a(w)=(m-1) \pmod{m}\}$. All range of values $\alpha\in
[1,m+1]$ can be reached for the nondeterministic state complexity of
the star of binary
languages~\cite{jiraskova08:_state_compl_of_compl_stars}.

\begin{figure}[htb]
  \begin{center}

\includegraphics[width=7cm]{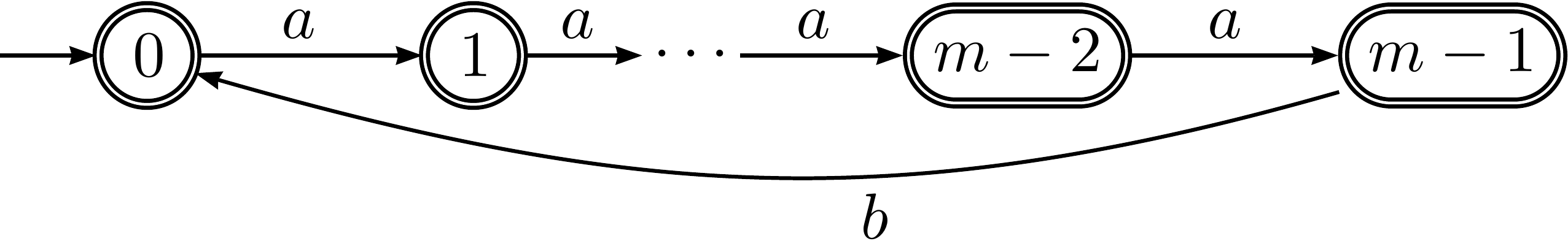}
\ignore{\SmallPicture{\VCDraw{
        \begin{VCPicture}{(-2,-2)(12,1)}
          \FinalState[0]{(0,0)}{0}\Initial{0}
          \FinalState[1]{(3,0)}{1}
          \FinalStateVar[m-2]{(8,0)}{m2}
          \SetStateLineStyle{none}
          \State[\cdots]{(5,0)}{2}
          \SetStateLineStyle{solid}
          \FinalStateVar[m-1]{(12,0)}{m1}
          \EdgeL{0}{1}{a}
          \EdgeL{1}{2}{a}
          \EdgeL{2}{m2}{a}
          \EdgeL{m2}{m1}{a}
          \VArcL{arcangle=16}{m1}{0}{b}
        \end{VCPicture}
      }
     }}

    \caption{\small Witness \nfas for the nondeterministic state complexity
      of reversal}
  \label{fig:jirasreveralnfa}
  \end{center}
\end{figure}

For the reversal, at most one more state will be
needed, so $nsc(L^R)\leq m+1$. Witness ternary languages were
presented by Holzer and Kutrib, but the bound is tight even for the family
of binary languages $(m>1)$ which minimal \nfas are presented in
Figure~\ref{fig:jirasreveralnfa}~\cite{jiraskova05:_state_compl_of_some_operat}.
If $nsc(L)=m\geq 3$ the possible values for  $nsc(L^R)$ are $m-1$, $m$ or
$m+1$~\cite{jiraskova08:_state_compl_of_compl_stars}. The first value
is reached by the reversals of the above binary languages and the
second considering the languages $\{w \in \{a,b\}^\star \mid \; |w|=0 \pmod{m}\}$.

The nondeterministic state complexity of left and right quotients by a
word were studied by
Ellul~\cite{ellul03:_descr_compl_measur_of_regul_languag}. Given a
minimal \nfa $A=(Q,\Sigma,\delta,q_0,F)$ accepting $L$, an \nfa $C$
accepting $Lw^{-1}$, for $w\in \Sigma$, coincides with $A$ except
that the set of final states is $\{q\in Q\mid \delta(q,w)\cap
F\not=\emptyset\}$. Thus $nsc(Lw^{-1})\leq nsc(L)$.  The witness
languages used for the state complexity of right quotient show that
the bound is tight. An upper bound for $nsc(w^{-1}L)$ can be obtained
by considering an \nfa $C$ with one new initial state $q_0'$ and
$\varepsilon$-transitions from $q_0'$ to each state of $A$ reached
when inputing $w$.

\begin{figure}[htb]
  \begin{center}
\includegraphics[width=8.5cm]{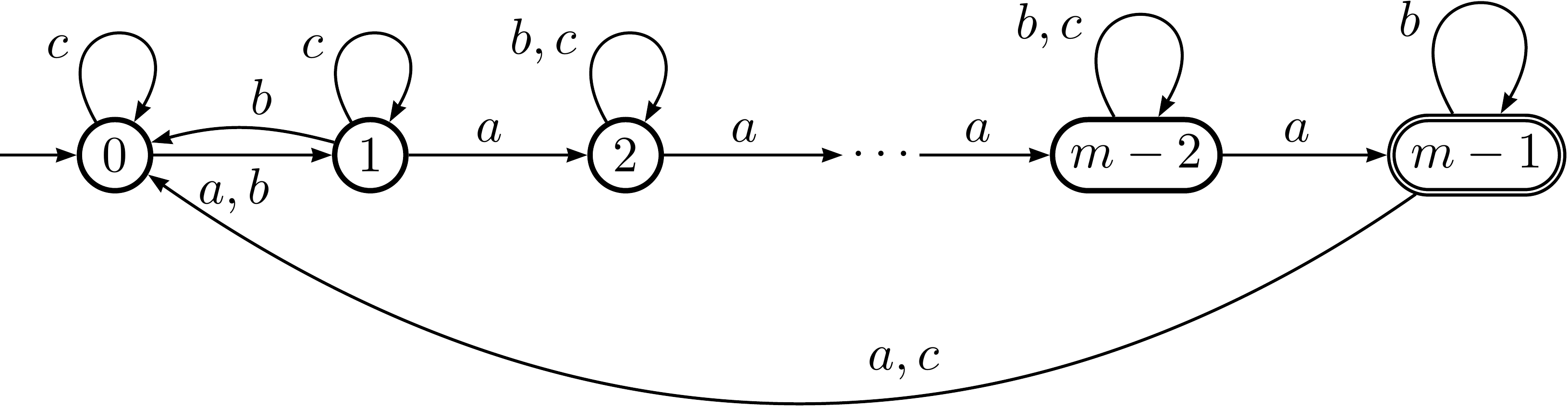}\ignore{\SmallPicture{\VCDraw{
      \begin{VCPicture}{(-2,-4)(15,2)}
      \State[0]{(0,0)}{0}\Initial{0}
      \State[1]{(3,0)}{1}
      \State[2]{(6,0)}{2}
      \SetStateLineStyle{none}
      \State[\cdots]{(9,0)}{3}
      \SetStateLineStyle{solid}
      \StateVar[m-2]{(12,0)}{m2}
      \FinalStateVar[m-1]{(16,0)}{m1}
      \EdgeR{0}{1}{a,b}
      \ArcR[.40]{1}{0}{b}
      \LoopN{0}{c}
      \LoopN{1}{c}
      \EdgeL{1}{2}{a}
      \LoopN{2}{b,c}
      \EdgeL{2}{3}{a}
      \EdgeL{m2}{m1}{a}
      \EdgeL{3}{m2}{a}
      \LoopN{m2}{b,c}
      \LoopN{m1}{b}
      \VArcR{arcangle=35}{m1}{0}{a,c}
    \end{VCPicture}
      }
     }}

    \caption{\small Universal witness \dfas,  $U_m(a,b,c)$.}

  \label{fig:universalwitness}
  \end{center}
\end{figure}

\paragraph{Universal Witnesses}
\label{sec:universalwitnesses}
Brzozowski~\cite{brzozowski12:_in_searc_of_most_compl_regul_languag,brzozowski13:_in_searc_of_most_compl_regul_languag}
identified a ternary family of languages $U_m(a,b,c)$ which provides witnesses for the state
complexity of all operations considered in the previous section.  The
family, presented in Figure~\ref{fig:universalwitness}, fulfills also
other conditions that, according to the same author, should be verified by
\emph{the most difficult (regular) languages}. For a language $L_m$
the suggested conditions are:
\begin{enumerate}[(1)]
\item \label{u:1} The state complexity should be $m$.
\item  \label{u:2} The state complexity of each quotient of $L_m$ should be $m$.
\item  \label{u:3} The number of atoms of $L_m$ should be $2^m$. An atom of a
  regular language with quotients $K_0,\ldots,K_{m-1}$ is a non-empty
  intersection of the form $\widetilde{K}_0\cap \cdots \cap \widetilde{K}_{m-1}$, 
where $\widetilde{K}_i$ is either $K_i$ or $\overline{K}_i$. Thus the
number of atoms is bounded from above by $2^m$, and it was proved by Brzozowski \textit{et al.}~\cite{brzozowski11:_theor_of_atomat,brzozowski14:_theor_of_atomat} that this bound is
tight\footnote{We also notice that the number of atoms of a language $L$ is equal to the state complexity of $L^{R}$.}. Every quotient of $L_m$  is a union of atoms.
\item  \label{u:4}The state complexity of each atom of $L_m$ should be maximal. 
It was shown~\cite{brzozowski12:_quotien_compl_of_atoms_of_regul_languag} that the  complexity of the atoms with 0 or $m$ complemented quotients is bounded from above by $2^m-1$, and the complexity of any atom with $r$ complemented quotients, where $1\le r\le m-1$, by 
\begin{equation*}
f(m,r)=1 + \sum_{k=1}^{r} \sum_{h=k+1}^{m-r+k} \binom{m}{h}\binom{h}{k}.
\end{equation*}
\item  \label{u:5} The syntactic semigroup  of $L_m$ should have cardinality 
  $m^m$, which is well known to be a tight upper
  bound~\cite{maslov70:_estim_of_number_of_states}. This measure,
  which is called the \emph{syntactic complexity} of a language, has been
  recently studied for many classes of subregular languages~\cite{
  holzer04:_deter_finit_autom_and_syntac_monoid_size,
  krawetz05:_state_compl_and_monoid_of,
  brzozowski12:_syntac_compl_of_prefix_suffix,
  brzozowski11:_syntac_compl_of_ideal_and_closed_languag,
  brzozowski12:_syntac_compl_of_some_class,
  brzozowski14:_syntactic_comp_trivial,  brzozowski14:_bounds_syntactic_compl}.
\end{enumerate}
The following result~\cite{brzozowski12:_in_searc_of_most_compl_regul_languag, brzozowski13:_in_searc_of_most_compl_regul_languag} can be considered a milestone in the operational state complexity for regular languages,  where $U_m$ is depicted in Figure \ref{fig:universalwitness}:

\begin{quotation}
  \emph{$(U_m(a,b,c)\mid m\geq 3)$ meets conditions \ref{u:1}--\ref{u:5} and
  is a witness for the reversal and the star. The families   $(U_m(a,b,c)\mid m\geq 3)$ and
    $(U_n(b,a,c)\mid n\geq 3)$ are witnesses for the Boolean
    operations, whereas $(U_m(a,b,c)\mid m\geq 3)$ and
    $(U_n(a,b,c)\mid n\geq 3)$ are witnesses for catenation.}
\end{quotation}

Variants of the universal witness were also given for several combined
operations. The question of whether there are universal witnesses for other
operations,  classes of subregular languages or other complexity
measures is an open problem (see~\cite{brzozowski14:_most_complex_right}). However, when searching for witnesses for
a given upper bound, to ensure that the above conditions (or some of
them) are verified, can be a good starting point. Moreover, the study of properties that may enforce  (some of) the conditions (\ref{u:1}) -- 
 (\ref{u:5}) is fundamental for a better understanding of the operational state complexity \cite{brzozowski14:_maxim_atomic_languag}.

\subsubsection{Unary Regular Languages}
\label{sec:unaryregularbop}
Table~\ref{tab:scnscbounary} presents the main state complexity
results of the basic operations on unary languages. Given the
constraints on both \dfas and \nfas over a one symbol alphabet, and
the results presented in Section~\ref{sec:scnsc}, the state complexity
for several operations on unary languages is much lower than what
is predicted by the general results of state complexity. Some results on the average-case state complexity of operations on
unary languages were presented by
Nicaud~\cite{nicaud99:_averag_state_compl_of_operat_unary_autom,nicaud00:_du_compor_en_moyen_des}.

\begin{table}[htbp]
  \centering
  \begin{tabular}{|l||c|c|c|}\hline
    \multicolumn{4}{|c|}{Unary Regular}\\\hline
    &sc&nsc&asc\\\hline\hline
    $L_1\cup L_2$&$\sim mn$ &
      $m+n+1$, if $m\not=\dot{n}$
&$\sim \displaystyle{\frac{3\zeta(3)}{2\pi^2}}mn$
    \\\hline
    $L_1\cap L_2$&$\sim mn$ &$mn$, if $(m,n)=1$&

$\sim \displaystyle{\frac{3\zeta(3)}{2\pi^2}}mn$
\\\hline 
    $\comp{L}$&$m$&$e^{\Theta(\sqrt{m\log m})}$&\\   \hline
    $L_1L_2$&$\sim mn$&\begin{tabular}{c}$[m+n-1,m+n]$,\\ if $m,n>1$
    \end{tabular}
    &$O(1)$, $n<P(m)$ 
\\\hline
    $L^\star$&
\begin{tabular}{c}$(m-1)^2+1$, \\if $m>1$, $l>1$
\end{tabular}
&$m+1$, if $m>2$
&$O(1)$
\\\hline 
    $L^{+}$&$(m-1)^2$&$m$, if $m>2$
&\\\hline
    $L^{R}$&$m$&$m$&\\\hline
    $w^{-1}L$&$m$&$m$&\\\hline
    $Lw^{-1}$&$m$&$m$&\\\hline
  \end{tabular}
  \caption{\small{State complexity ($sc$), nondeterministic state complexity ($nsc$) and
      average state complexity ($asc$) of
      basic operations on unary languages. The $\sim$ symbol means that the complexities are asymptotically equal to the given values. The upper bounds of state
      complexity for union, intersection and catenation  are exact if the greatest common divisor of $m$ and $n$,      $(m,n)$ is $1$. For the average state complexity of intersection and
      union, $\zeta(n)$ is the function $\zeta$ of Riemman. 
For the average state complexity of catenation, $n$ must
      be bounded by a polynomial $P$ in $m$. 
    }
  }
  \label{tab:scnscbounary}
\end{table}
A \dfa that accepts a unary language is characterized by a noncyclic
part (the tail) and a cyclic part (the loop).  A characterization and
the enumeration of minimal unary \dfas was given by
Nicaud~\cite{nicaud99:_averag_state_compl_of_operat_unary_autom}.

The state complexity of the reversal of a unary language $L$ is
trivially equal to the state complexity of $L$.  The state complexities of Boolean operations on unary languages coincide asymptotically with the ones on
 general regular
languages. Yu~\cite{yu01:_state_compl_of_regul_languag} shown that the
bound was tight for union (and thus, for intersection) if $m$ and $n$
are coprimes and the witness languages are $(a^m)^\star$ and
$(a^n)^\star$. The state complexity of catenation and star was
proved by Yu \emph{et al.}~\cite{yu94:_state_compl_of_some_basic} and
the tightness for the first was also shown for $m$ and $n$
coprimes. The witnesses for the catenation are $(a^m)^\star a^{m-1}$ and $(a^n)^\star a^{n-1}$. For the star, if $m=2$ a
witness is $(aa)^\star$, and for each $m> 2$ a witness is $(a^m)^\star
a^{m-1}$.  The state complexity when $m$ and $n$ are not necessarily
coprimes was studied by Pighizzini and
Shallit~\cite{pighizzini01:_unary_languag_concat_and_its_state_compl,pighizzini02:_unary_languag_operat_state_compl}. In
this case, the tight bounds are given by the number of states in the
tail and in the loop of the resulting automata.  The state complexity
for left and right quotient by a word on unary languages coincide
with the general case.

 Nicaud~\cite{nicaud99:_averag_state_compl_of_operat_unary_autom,nicaud00:_du_compor_en_moyen_des}
 proved that the state complexity of union, intersection and
 catenation on two languages $L_1$ and $L_2$ is asymptotically
 equivalent to $mn$, where $m=sc(L_1)$ and $n=sc(L_2)$.  Let $D_n$  be  the set  of
 unary (complete and initially connected) \dfas
 with $n$ states. The \emph{average state complexity}
 (asc) of a
 binary operation $\circ$ on regular languages is given by 
$$\frac{\displaystyle{\sum_{A_1\times A_2\in D_m\times D_n}}sc(L(A_1)\circ
  L(A_2))}{|D_m\times D_n|}$$

\noindent This definition can be generalized to  operations with other 
arities, other kinds of automata and classes of languages.

As shown in Table~\ref{tab:scnscbounary}, the average state complexities 
of catenation and star  on unary languages are bounded by a
constant, and for intersection (and union)  note that 
$\frac{3\zeta(3)}{2\pi^2}\approx 0.1826907423$.
 Magical numbers for the star operation on unary languages was studied by \v Cevorová~\cite{cevorova13}. Considering the gap between the worst-case upper bound, $n^2-2n+2$, and the average case (less than a constant), it is not a surprise that for every $n$ no more than $4$ complexities are attainable between $n^2-4n+6$ and the upper bound. In the same paper, the author also establishes a relation between this problem and the Frobenius problem.

The nondeterministic state complexity of basic operations on unary
languages was studied by Holzer and
Kutrib~\cite{holzer03:_unary_languag_operat_and_their}, and also by
Ellul~\cite{ellul03:_descr_compl_measur_of_regul_languag}. For union
and intersection, the upper bound coincides with the general
case. However, it was proved to be achievable for union if $m$ is not
a divisor or multiple of $n$. As in the deterministic case, the
witnesses for intersection are $(a^m)^\star$ and $(a^n)^\star$, if
$m$ and $n$ are coprimes. The nondeterministic state complexity of the
complementation is $O(F(m))$ (where $F$ is the Landau's function
of equation (\ref{eq:fm})),
which is directly related with the state complexity of
determination. Holzer and
Kutrib~\cite{holzer03:_unary_languag_operat_and_their} proved that
this upper bound is tight in order of magnitude, i.e.,  for any integer $m > 1$ there exists a unary language $L$ such that
 $nsc(L)=m$ and $nsc(\comp{L})=\Omega(F(m))$.
Moreover, Mera and Pighizzini~\cite{mera05:_compl_unary_nondet_autom}
have shown that for each $m\geq 1$ and unary language $L$, such that
$nsc(L)=m$ and $sc(L)=sc(\comp{L})=e^{O(\sqrt{m\log m})}$, then
$nsc(\comp{L})\geq m$. The upper bound $m+n$ for the catenation of two unary languages is not
know to be tight. The known lower bound is $m+n-1$ achieved by the
catenation of  $\{a^l\mid l= (m-1) \pmod{m}\}$ and $\{a^l\mid l= (n-1)
\pmod{n}\}$~\cite{holzer03:_unary_languag_operat_and_their}.
The same languages can be used to show the tightness of the bound
$m+1$ for the star (and the plus) operation. For the left and right quotients, notice that in the unary case
$w^{-1}L=Lw^{-1}$, and the results for the general case apply.

\subsubsection{Finite Languages}
\label{sec:finitebop}
Finite languages are an important subset of regular languages. They
are accepted by complete \dfas that are acyclic apart from a loop on
the \emph{sink} (or dead) state, for all alphabetic symbols. Minimal
\dfas have also special graph properties that lead to a linear time
minimization algorithm~\cite{revuz92:_minim_of_acycl_deter_autom}, and where
the length of the longest word accepted by the language plays an important role. Table~\ref{tab:scncsfiniteregular} shows that the (nondeterministic)
state complexity of operations on finite languages are, in general,
lower than in the general case.

\begin{figure}[htb]
  \centering
\begin{tabular}[c]{c}

  \raisebox{-.5\height}{\includegraphics[width=6cm]{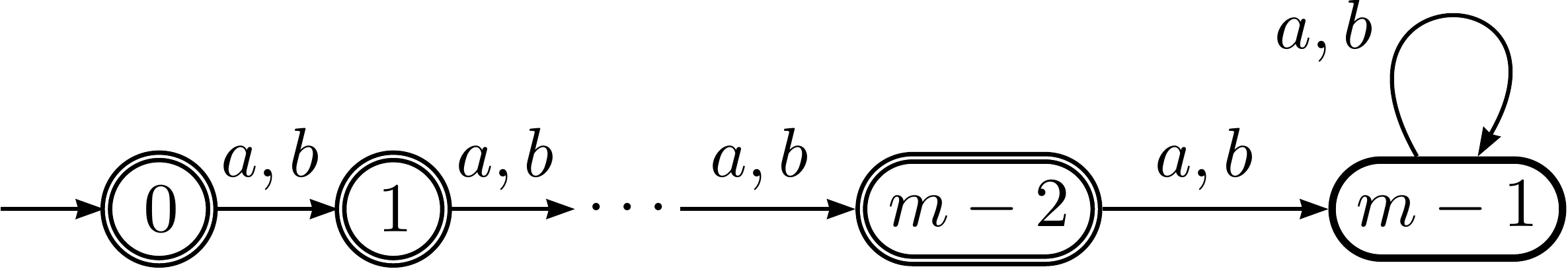}}
	\ignore{\SmallPicture{\VCDraw{
        \begin{VCPicture}{(-1,-1)(11,3)}
          \FinalState[0]{(0,0)}{0}\Initial{0}
          \FinalState[1]{(2,0)}{1}
          \SetStateLineStyle{none}
          \State[\cdots]{(4,0)}{2}
          \SetStateLineStyle{solid}
          \FinalStateVar[m-2]{(7,0)}{m2}
          \StateVar[m-1]{(11,0)}{m1}
          \EdgeL{0}{1}{a,b}
          \EdgeL{1}{2}{a,b}
          \EdgeL{2}{m2}{a,b}
          \EdgeL{m2}{m1}{a,b}
          \LoopN{m1}{a,b}
         \end{VCPicture}
       }
     }}
  \\\raisebox{-.5\height}{\includegraphics[width=6cm]{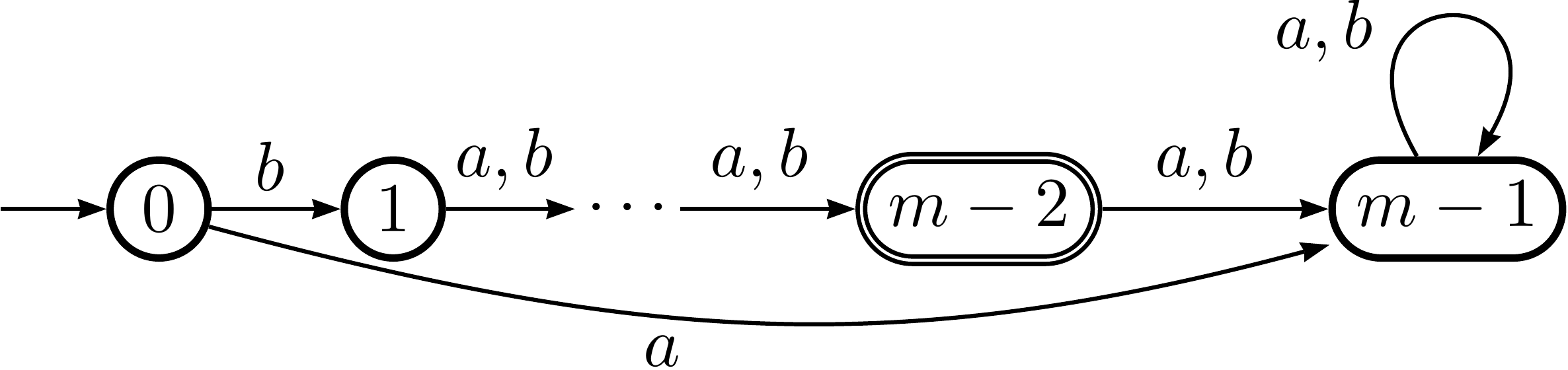}}
	\ignore{\SmallPicture{\VCDraw{
        \begin{VCPicture}{(-1,-1)(11,3)}
          \State[0]{(0,0)}{0}\Initial{0}
          \State[1]{(2,0)}{1}
          \SetStateLineStyle{none}
          \State[\cdots]{(4,0)}{2}
          \SetStateLineStyle{solid}
          \FinalStateVar[m-2]{(7,0)}{m2}
          \StateVar[m-1]{(11,0)}{m1}
          \EdgeL{0}{1}{b}\ArcR{0}{m1}{a}
          \EdgeL{1}{2}{a,b}
          \EdgeL{2}{m2}{a,b}
          \EdgeL{m2}{m1}{a,b}
          \LoopN{m1}{a,b}
        \end{VCPicture}
      }
    }}
  \end{tabular}    
\caption{\small{Witness \dfas for the state complexity  of catenation on
  finite languages}}
\label{fig:finiteintersection}
\end{figure}

\begin{table} [htbp]
  \centering
\begin{tabular}{|l||c|c|c|c|}\hline
\multicolumn{5}{|c|}{Finite}\\\hline
&\multicolumn{1}{c}{sc}&\multicolumn{1}{c|}{$|\Sigma|$}&\multicolumn{1}{c}{nsc}&\multicolumn{1}{c|}{$|\Sigma|$}\\\hline\hline
$L_1\cup L_2$&$mn-(m+n)$&$f(m,n)$&$m+n-2$&2\\\hline
$L_1\cap L_2$&$mn - 3(m + n) + 12$&$f(m,n)$&$mn$&2\\\hline
$\comp{L}$&$m$&1&$\Theta(k^{\frac{m}{1+\log k}})$&2\\\hline
\multirow{2}{*}{$L_1L_2$}&$(m-n+3)2^{n-2}-1$, $m+1\geq n$&$2$&\multirow{2}{*}{$m+n-1$}&\multirow{2}{*}{$2$}\\\cline{2-3}
&$m+n-2$, if $l_1=1$&$1$&&\\\hline
\multirow{2}{*}{$L^\star$}&
$2^{m-3}+2^{m-l-2}$, $l\geq 2$, $m\geq 4$&3&\multirow{2}{*}{$m-1$, $m>1$}&\multirow{2}{*}{$1$}\\\cline{2-3}
&$m-1$, if $f=1$&$1$&&\\\hline
$L^{+}$&$m$&$1$&$m$, $m>1$&1\\\hline
$L^{R}$&$O(k^{\frac{m}{1+\log k}})$&2&$m$&2\\\hline
\end{tabular}
\caption{\small{State complexity and nondeterministic state complexity of
  basic operations on finite languages}}
\label{tab:scncsfiniteregular}
\end{table}
  C\^ampeanu \emph{et
  al.}~\cite{campeanu01:_state_compl_of_basic_operat_finit_languag}
presented the first formal study of state complexity of operations on
finite languages. Yu~\cite{yu01:_state_compl_of_regul_languag}
presented upper bounds of $O(mn)$ for the union and the intersection.
The tight upper bounds were given by Han and
Salomaa~\cite{han08:_state_compl_of_union_and} using growing size
alphabets. The upper bound for union and intersection cannot be
reached with a fixed alphabet when $m$ and $n$ are arbitrarily large.
C\^ampeanu \emph{et al.} gave tight upper bounds for catenation, star
and reversal. For catenation the bound $(m-n+3)2^{n-2}-1$ is tight for
binary languages, if $m+1\geq n> 2$. The \dfas of the witness
languages are presented in Figure~\ref{fig:finiteintersection}.

For star, C\^ampeanu \emph{et al.}  shown that the bound
$2^{m-3}+2^{m-4}$ is tight for ternary languages.  The tight upper
bound for the reversal of a finite binary language is
$3\cdot2^{p-1}-1$, if $m=2p$, and $2^{p-1}-1$ if $m=2p-1$.

Nondeterministic state complexity of basic operations on finite
languages were studied by Holzer and
Kutrib~\cite{holzer03:_state_compl_of_basic_operat}.  Minimal \nfas
accepting finite languages without the empty word can be assumed to
have only a final state (with no transitions); and if the empty word is in
the language, the initial state is also final.  Because there
are no cycles, for the union of two finite languages three states can
be avoided: no new initial state is needed, and the initial states and
the final states can be merged. The upper bound $m+n-2$ is tight for
the languages ${a^{m-1}}$ and ${b^{n-1}}$, for $m,n\geq 2$.  In the case
of the intersection, the upper bound coincides with the general case,
and it is tight for the binary languages $\{w\in\{a,b\}^\star \mid
\#_a(w)=0 \pmod{m}\}$ and $\{w\in\{a,b\}^\star \mid \#_b(w) = 0
\pmod{n}\}$. Considering the upper bound of determination for finite
languages, the nondeterministic state complexity for complement is
bounded by $O(k^{\frac{m}{1+\log k}})$. The lower bound
$\Omega(k^{\frac{m}{2\log k}})$ is reached for alphabets
$\Sigma=\{a_1,\ldots,a_k\}$ of size $k\geq 2$, and the languages
$\Sigma^ja_1\Sigma^iy$, where $i\geq 0$, $0 \leq j \leq i$, $y \in
\Sigma\setminus\{a_1\}$, and $m>2$. However, a tighter lower bound 
can be achieved by the determination lower bound of $\Omega(k^{\frac{m}{1+\log k}})$.
 For catenation of finite languages represented by \nfas, one state
can be saved. Witness languages for the tightness of the bound $m+n-1$
can be the ones used for union. Two states are also saved for the
star, and for plus the nondeterministic state complexity coincides with
the one for the general case. Witness languages are $a^{m}$
and $a^{m-1}$, respectively. \nfas for the reversal are exponentially more succinct then \dfas. In
the case of finite languages, and like other operations, one 
state can be spared. Witness languages are $\{a,b\}^{m-1}$.

\subsubsection{Finite Unary Languages}
\label{sec:scnscfiniteunary}
Table~\ref{tab:scncsfiniteunary} summarizes the state complexity and
nondeterministic state complexity results of basic operations on
finite unary
languages~\cite{campeanu01:_state_compl_of_basic_operat_finit_languag,yu01:_state_compl_of_regul_languag,holzer03:_unary_languag_operat_and_their}.
State complexity of union, intersection and catenation on finite unary
languages are linear, while they are quadratic for general unary
languages. In this setting, nondeterminism is only relevant for the
star (and plus), as unary regular languages are obtained.  As 
already stated, for a finite unary language $L$, one has $sc(L)\leq
nsc(L)+1$, and $sc(L)-2$ is the length of the longest word in the
language. If a operation preserves finiteness, for state complexity
only the longest words must be considered.

\begin{table}[htbp]
  \centering
\begin{tabular}{|l||c|c|}\hline
\multicolumn{3}{|c|}{Finite Unary}\\\hline
&sc&nsc\\\hline\hline
$L_1\cup L_2$&$\max\{m,n\}$&$\max\{m,n\}$\\\hline

$L_1\cap L_2$&$\min\{m,n\}$
&$\min\{m,n\}$
\\\hline
$\comp{L}$&$m$
  &$m+1$\\\hline
  $(L_1- L_2)$&$m$ & \\\hline 
  $(L_1\oplus L_2)$&$\max\{m,n\}$&\\\hline
$L_1L_2$&$m+n-2$&$m+n-1$\\\hline
$L^\star$&
\begin{tabular}{c}
$2$, if $m=3$\\
$m-1$, if $f=1$\\
$m^2-7m+13$, if $m>4$, $f\geq 3$
\end{tabular}&
$m-1$\\\hline
$L^{+}$&$m$&$m$\\\hline
$L^{R}$&$m$&$m$\\\hline
\end{tabular}
\caption{\small{State complexity and nondeterministic state complexity of
  basic operations on finite unary languages}}
\label{tab:scncsfiniteunary}
\end{table}

\subsection{Other Regularity Preserving Operations}
\label{sec:otherpreservingregularityoperations}

Table~\ref{tab:scnscregularitypreservingoperations} presents the
results for the state complexity of some regularity preserving
operations, that are detailed in the next paragraphs.

\paragraph{Proportional removals}
\label{sec:scnscpropotionalremovals}
The \emph{proportional removals} preserving
regularity were studied by
Hartmamis~\cite{stearns63:_regul_preser_modif_of_regul_expres} and
were full characterized by
Seiferas and McNaughton~\cite{seiferas76:_regul_preser_relat}.
 For any binary relation $r \subseteq \mathbb{N} \times \mathbb{N} $ 
and any language $L \subseteq \Sigma^\star$, let the language $P (r, L)$ be defined as
$$P(r,L)=\{x\in\Sigma^\star\mid \exists y\in\Sigma^\star \text{ such
  that } xy\in L \,\wedge\, r(|x|,|y|)  \}.$$  

\noindent A relation $r$ is
\emph{regularity-preserving} if $P(r,L)$ is regular for every regular
language $L$. Seiferas and
McNaughton~\cite{seiferas76:_regul_preser_relat} gave sufficient and
necessary conditions of regularity preservation in this context.  For the special case where $r$ is the identity, the correspondent
language is denoted by $\frac{1}{2}(L)$.
Domaratzki~\cite{domaratzki02:_state_compl_of_propor_remov} proved
that for a regular language $L$,
$sc(\frac{1}{2}(L))=O(sc(L)F(sc(L)))$ (where $F$ is the Landau's
function of equation (\ref{eq:fm})) and this bound is tight for ternary languages. In the
case of $L$ be a unary language, one gets $sc(\frac{1}{2}(L))=sc(L)$.
Following Nicaud's work on average-case complexity, mentioned above,
Domaratzki showed that the average state complexity of the
$\frac{1}{2}(\cdot)$ operation on a $m$-state unary automaton is
asymptotically equivalent to $\frac{5}{8} m + c$, for some constant
$c$. Domaratzki also studied the state complexity of polynomial removals.  Let $f \in \mathbb{Z}[x]$ be a strictly monotonic polynomial such
   that $f(\mathbb{N}) \subset \mathbb{N}$. Then, the relation $r_f =
   \{(n, f (n))\mid n \geq 0\}$ preserves regularity, and $sc(P
   (r_f,L)) \leq O(sc(L)F(sc(L))).$


\noindent In 1970, Maslov~\cite{maslov70:_estim_of_number_of_states} had already
studied the language $\frac{p}{q}(L)$, i.e., a language $P(r,L)$ such
that $r$ is defined by $\{(m,n)\mid mq=pn \}$ with $p, q\in
\mathbb{N}$.  An open problem is to obtain the state complexity of
$P(r,L)$ where $r$ belongs to the broader class of regularity preserving
relations studied by Seiferas and McNaughton. 

Nondeterministic state complexity of polynomial removals was studied by Go\v c \emph{et al.}~\cite{GocPS13}. The authors showed an $O(n^2)$ upper bound and a matching lower bound in the case where the polynomial is a sum of monomials and a constant, or when the polynomial has rational roots.

\begin{table}[htbp]
  \centering
\begin{tabular}{|c||c|c|c|c|}\hline
\multicolumn{5}{|c|}{Regular}\\\hline
&\multicolumn{1}{c}{sc}&$|\Sigma|$&\multicolumn{1}{c}{nsc}&$|\Sigma|$\\\hline\hline
\multirow{2}{*}{$\frac{1}{2}(L)$}& $me^{\Theta(\sqrt{m\log m})}$&3&\multirow{2}{*}{$O(m^2)$}&\\\cline{2-3}
&$m$&1&&\\\hline

\multirow{2}{*}{$L^i$}&$\Theta(m2^{(i-1)m})$&$6$&\multirow{3}{*}{$im$}&\multirow{3}{*}{$2$}\\\cline{2-3}
&$im-i+1$&$1$&&\\\cline{1-3}
$L^3$&$\frac{6m-3}{8}\scriptstyle{4^m-(m-1)2^m-m}$&$4$&&\\\hline
\multirow{3}{*}{$\shift{L}$}&$2^{m^2+m\log m-O(m)}$&$4$&$1$, if $m=1$&\multirow{2}{*}{$2$}\\\cline{2-4}
&$2^{\Theta(m^2)}$& $2$,$3$&$2m^2+1$, if $m\geq 2$&\\\cline{2-5}
&$m$&$1$&$m$&$1$\\\hline
$L_1\shuffle L_2$&$O(2^{mn}-1)$
&$5$&$O(mn)$
&$5$\\\hline
\multirow{2}{*}{$L_1\odot_{\bot}L_2$}&$m2^{n-1}-2^{n-2}$,&\multirow{2}{*}{$4$}&
\multirow{2}{*}{$m+n$}&\multirow{2}{*}{$2$}\\
&if $m\geq 3$, $n\geq  4$&&&\\\hline
\multicolumn{5}{|c|}{Unique Regular Operations}\\\hline
$L_1\stackrel{\circ}{\cup}L_2$&$mn$&$2$&&\\\hline
$L_1\circ L_2$&$O(m3^n-f_13^{n-1})$&&$\geq 2^{O(h)}$&\\\hline
${L}^{\circ 2}$&$m3^m-3^{m-1}$&$2$&&\\\hline
${L}^{\circ}$&
$\begin{array}{c}
  O(3^{m-1} + (f+2)3^{m-f-1}\\
  - (2^{m-1} + 2^{m-f-1}-2))
\end{array}$
&&&\\\hline
\end{tabular}
\caption{\small State complexity and nondeterministic state complexity of
  some regularity preserving operations: proportional removals for
  the identity relation ($\frac{1}{2}(L))$; power $L^i$ where
  $i\geq 2$; cyclic shift $\shift{L}$; shuffle $L_1\shuffle L_2$;
 orthogonal catenation
  $L_1\odot_{\bot}L_2$; unique operations: for unique star $L^{\circ}$,
  $\varepsilon\notin L$; for the nondeterministic state complexity of $L_1\circ L_2$,
  the combined state complexity of $L_1$ and $L_2$ is $O(h)$, for $h\geq 0$.}
 \label{tab:scnscregularitypreservingoperations}
\end{table}

\paragraph{Power}
\label{sec:scnscpower}

Given a regular language $L$ and $i\geq 2$, an upper bound of the
state complexity of the language $L^i$ is given by considering the state
complexity of catenation. However, a tight upper bound is obtained if
this operation is studied individually.  Domaratzki and
Okhotin~\cite{domaratzki09:_state_compl_of_power} proved that
$sc(L^i)=\Theta(m2^{(i-1)m})$, for $i\geq 2$.  
The bound is tight
for a family of languages over a six-symbol alphabet. In the case $i =
3$, $sc(L^3)=\frac{6m - 3}{8} 4^m - (m -1)2^m - m$, for $m\geq 3$, and the
tightness is witnessed by a family of languages over a four-symbol
alphabet. For the square, i.e. if $i=2$, the upper bound is the one given by the state complexity of catenation, $sc(L^2)=m2^m -2^{m-1}$ and it is met by a language accepted by a $m$-state DFA with only one final state. In the case of multiple $l$ final states, the upper bound is $(m-l)2^m+l2^{m-1}$. \v Cevorová \emph{et al.}~\cite{CevorovaJK14} proved that those upper bounds are tight in the ternary case for every $l\in[1,m-2]$. The nondeterministic state complexity of $L^i$ is
proved to be $im$. This bound is shown to be tight over a binary
alphabet, for $m\geq 2$.
The power of unary languages was studied by
Rampersad~\cite{rampersad06:_state_compl_of_l_and_l_k}. If $L$ is a
unary language with $sc(L)=m\geq 2$, then $sc(L^i)=im-i+1$. For the square,  \v Cevorová \emph{et al.} showed that all the complexities in the range $[1,2m-1]$ can be
attained for $m\geq 5$.

\paragraph{Cyclic Shift}
\label{sec:scnsccyclicshift}


The \emph{cyclic shift} of a language $L$ is defined as $\shift{L} = \{
vu \mid uv \in L \}$.
Maslov~\cite{maslov70:_estim_of_number_of_states} gave an upper bound
of $(m2^m-2^{m-1})^m$ for the state complexity of cyclic shift and an
asymptotic lower bound of $(m-3)^{m-3} \cdot 2^{(m-3)^2}$, by
considering languages over a growing alphabet (if complete \dfas
are considered). Jir\'askov\'a and
Okhotin~\cite{jiraskova08:_state_compl_of_cyclic_shift} reviewed and
improved Maslov results. Using a fixed four-symbol alphabet, they
obtained a lower bound of $(m-1)! \cdot 2^{(m-1)(m-2)}$, $m\geq 3$,  which shows
that $sc(\shift{L})=2^{m^2 +m\log m - O(m)}$
for alphabets of size greater than $3$. For binary and ternary languages, they
proved that the state complexity is $2^{\Theta(m^2)}$. 
As this function grows faster than the number
of \dfas for a given $m$, there must exist some \emph{magic numbers}
for the  state complexity of the cyclic shift over languages of a fixed alphabet. 

The nondeterministic state complexity of this operation was shown to
be $2^{m^2}+1$, for $m\geq 2$, and the upper bound is tight for binary
languages. Although the hardness of this operation on the
deterministic case, its  nondeterministic state complexity is
relatively low.  For a unary language $L$, as $\shift{L}=L$, one
gets $sc(\shift{L})=nsc(\shift{L})=sc(L)$.

\paragraph{Shuffle}\label{sec:scnscshuffle}

\begin{figure}[ht]
  \centering

  \begin{tabular}[t]{c}
    \raisebox{-.5\height}{\includegraphics[width=6.5cm]{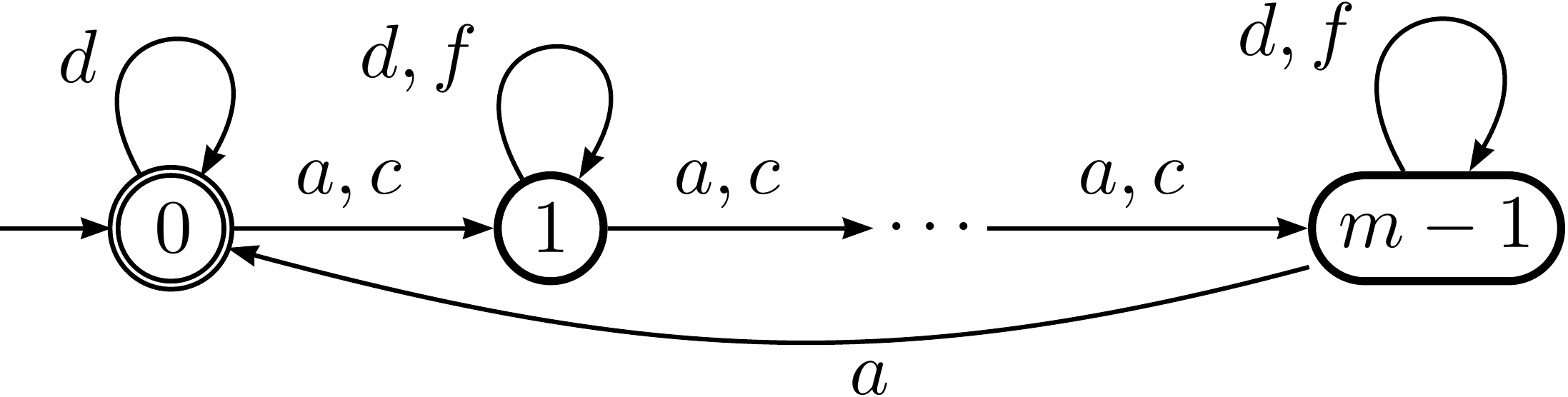}}
		\ignore{\SmallPicture{\VCDraw{
            \begin{VCPicture}{(-2,-2)(11,2)}
              \FinalState[0]{(0,0)}{0}\Initial{0}
              \State[1]{(3,0)}{1}
              \SetStateLineStyle{none}
              \State[\cdots]{(6,0)}{4}
              \SetStateLineStyle{solid}
              \StateVar[m-1]{(10,0)}{m1}
              \LoopN{0}{d}
              \LoopN{1}{d,f}
              \LoopN{m1}{d,f}
              \EdgeL{0}{1}{a,c}
              \EdgeL{1}{4}{a,c}
              \EdgeL{4}{m1}{a,c}
              \ArcL{m1}{0}{a}
            \end{VCPicture}
          }
        }}
    \\\raisebox{-.5\height}{\includegraphics[width=6.5cm]{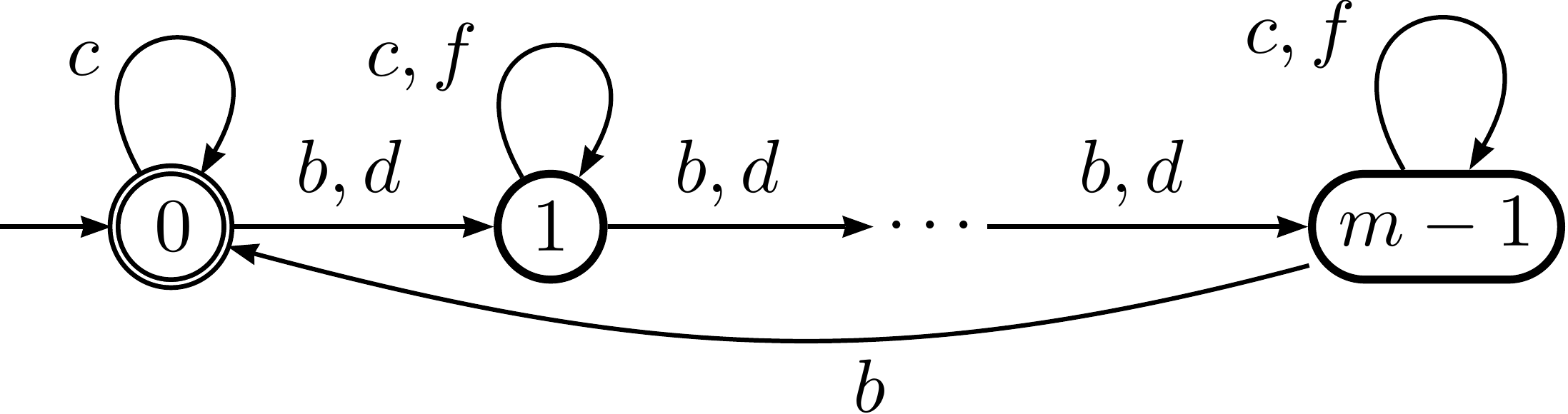}}
		\ignore{\SmallPicture{\VCDraw{
            \begin{VCPicture}{(-2,-2)(11,2)}
              \FinalState[0]{(0,0)}{0}\Initial{0}
              \State[1]{(3,0)}{1}
              \SetStateLineStyle{none}
              \State[\cdots]{(6,0)}{4}
              \SetStateLineStyle{solid}
              \StateVar[m-1]{(10,0)}{m1}
              \LoopN{0}{c}
              \LoopN{1}{c,f}
              \LoopN{m1}{c,f}
              \EdgeL{0}{1}{b,d}
              \EdgeL{1}{4}{b,d}
              \EdgeL{4}{m1}{b,d}
              \ArcL{m1}{0}{b}
             \end{VCPicture}
           }
         }}

  \end{tabular}
  \caption{\small Incomplete \dfas for the tight upper bound of state
    complexity of shuffle.}
\label{fig:shufflewitnesses}
\end{figure}

The \emph{shuffle} operation  of two words $w_1,w_2\in \Sigma^\star$ is
defined by
\begin{eqnarray*}
  w_1\shuffle w_2&=&\{u_1 v_1\ldots u_mv_m\mid\\
&& u_i,v_i\in\Sigma^\star,\  i
\in[1,m],\, w_1=u_1\ldots u_m \text{ and } w_2=v_1\ldots v_m \}.
\end{eqnarray*}
This operation is extended trivially to languages. If two regular
languages are regular, their shuffle is also a regular language.
C\^ampeanu \emph{et al.}~\cite{campeanu02:_tight_lower_bound_for_state}
showed that the state complexity of the shuffle of two regular languages
$L_1$ and $L_2$ is less or equal to $2^{mn}-1$.
They proved that this bound is tight for witness
languages over a five symbols alphabet and if minimal incomplete
\dfas are considered (see Figure~\ref{fig:shufflewitnesses}). Thus,
$sc(L_1\shuffle L_2)$ is at least $2^{(sc(L_1)-1)(sc(L_2)-1)}$.

Various restrictions and generalizations of the shuffle operation have
been studied. Mateescu \emph{et al.}~\cite{mateescu98:_shuff_trajec}
introduced the shuffle operation of two languages $L_1$ and $L_2$ on a
set of trajectories $T\subseteq \{0,1\}^\star$, $L_1\shuffle_T L_2$.
When $L_1$, $L_2$, and $T$ are regular languages $L_1\shuffle_T L_2$
is a regular language. In particular, if $T=\{0,1\}^\star$, then
$L_1\shuffle_T L_2=L_1\shuffle L_2$; and if
$T=\{0\}^\star\{1\}^\star$, then $L_1\shuffle_T L_2=L_1L_2$.
Domaratzki and
Salomaa~\cite{domaratzki04:_state_compl_of_shuff_trajeccd} studied the
state complexity of the shuffle on regular trajectories. In general,
$sc(L_1\shuffle_T L_2)\leq 2^{nsc(L_1)nsc(L_2)nsc(T)}$. If $T$ belongs
to special families of regular languages, tight bounds were also
presented.

\paragraph{Orthogonal Catenation}
\label{sec:scnscorthogonalcatenation}

A language $L$ is the \emph{orthogonal catenation} of $L_1$ and $L_2$,
and denoted by $L = L_1 \odot_\bot L_2$, if every word $w$ of $L$ can
be obtained in just one way as a catenation of a word of $L_1$ and a
word of $L_2$. If catenation uniqueness is not verified for every word
of $L$, orthogonal catenation of $L_1$ and $L_2$ is undefined,
otherwise $L_1$ and $L_2$ are \emph{orthogonal}. Daley \emph{et
  al.}~\cite{daley10:_orthog_concat}
studied the state complexity of orthogonal catenation and generalized
orthogonality to other operations.  Although it is a restricted
operation, its state complexity is only half of the one for the
general catenation, i.e., $m2^{n-1}-2^{n-2}$ for $m\geq 3$ and $n\geq
4$. The tight bound was obtained for languages over a four-symbol
alphabet.  Concerning nondeterministic state complexity, one has
$nsc(L_1\odot_\bot L_2)=nsc(L_1)+nsc(L_2)$, which coincides with the
one for (general) catenation. Witness languages presented for the
catenation are orthogonal (see page \pageref{witnessnsccatenation}),
thus apply to orthogonal catenation.

\paragraph{Unique Regular Operations}
\label{sec:scnscuniqueoperations}

Similar to orthogonality is the concept of \emph{unique operation}
introduced by Rampersad \emph{et
  al.}~\cite{rampersad09:_state_compl_of_unique_ration_operat}. However,
instead of demanding that every pair of words of the operand languages
lead to a distinct word on the resulting language, the language
resulting from a \emph{unique operation} only contains the words that
are uniquely obtained through the given operation.  Rampersad \emph{et
  al.} studied
several properties of unique operations and of their \emph{poly}
counterpart (i.e. where each resulting word must be obtained in more
than one way), such as closure, ambiguity, and  membership and
non-emptiness decision problems. Results on
state complexity and nondeterministic state complexity were obtained
for \emph{unique union} ($L_1\stackrel{\circ}{\cup}L_2$),
 \emph{unique catenation} ($L_1\circ L_2$), \emph{unique
  square} ($L\circ L={L}^{\circ 2}$), and \emph{unique star} (${L}^\circ$).
The state complexity of $L_1\stackrel{\circ}{\cup}L_2$ is
$mn$, and witness binary languages are $\{x\in\{a,b\} \mid
\#_a(x) = (m-1)\pmod{m}\}$ and $\{x\in\{a,b\} \mid \#_b(x)= (n-1)
\pmod{n}\}$, for $m,n\geq 3$ (that were also used by Maslov~\cite{maslov70:_estim_of_number_of_states} for general union).
For unique catenation, $sc(L_1\circ L_2)\leq m3^n-f_13^{n-1}$ which is
much higher than the one for general catenation. It is an open problem
to know if this bound is tight, although  several examples, for specific values of
$m$ and $n$, were presented.  However, for the unique square
$sc({L}^{\circ 2})=m3^m-3^{m-1}$, and the bound is
tight for binary languages and $m\geq 3$.  For the nondeterministic state
complexity of unique catenation, a  exponential lower
bound was provided.
An upper bound for the state complexity of the unique star is
$3^{m-1} + (f+2)3^{m-f-1} - (2^{m-1} + 2^{m-f-1}-2)$. But,
again,  it is an open problem to know if this upper bound is tight.

\subsection{Other Subregular Languages}
\label{sec:subregularlanguages}

Besides finite and unary languages, several other subregular languages
are used in many applications and are now theoretically well studied.
Prefix-free or suffix-free languages are examples of codes that are
fundamental in coding
theory~\cite{jurgensen97:_codes,berstel10:_codes_and_autom}. Prefix-closed,
factor-closed, or subword-closed languages were studied by several
authors~\cite{haines69:_free_monoid_partial_order_by_embed,thierrin72:_convex_languag,luca90:_some_combin_proper_of_factor_languag,gill74:_multip_entry_finit_autom}. These
languages belong to a boarder set of languages, the \emph{convex
  languages}, for which a general framework have
been recently
addressed by Ang and
Brzozowski~\cite{ang09:_languag_convex_with_respec_to} and Brzozowski
\emph{et al.}~\cite{brzozowski09:_decis_probl_for_convex_languag}. A
detailed survey on complexity topics was presented
by Brzozowski~\cite{brzozowski10:_compl_in_convex_languag}.  Partially
based on that survey, here we summarize some of the results concerning
the state complexity of preserving regularity operations over some of
the convex subregular languages. Star-free languages are other family of 
subregular languages well studied~\cite{schutzenberger65:_finit_monoid_havin_only_trivial_subgr,mcnaughton71:_count_free_autom}. We
 briefly address recent results on the (nondeterministic) state complexity of basic
regular operations on these languages.

\subsubsection{Convex Subregular Languages}
\label{sec:convexlanguages}
We begin by some definitions and results
on determination for these languages. Let $\unlhd$ be a partial order on $\Sigma^\star$, and let $\unrhd$ be
its converse. A language $L$ is $\unlhd$-convex if $u \unlhd v$ and $v
\unlhd w$ with $u,w \in L$ implies $v \in L$. It is $\unlhd$-free if
$v \unlhd w$ and $w \in L$ implies $v \notin L$. It is $\unlhd$-closed if
$v \unlhd w$ and $w \in L$ implies $v \in L$. 
It is $\unrhd$-closed if $v \unrhd w$ and $w \in L$ implies $v \in L$.  
The closure and the
converse closure operations are:
$$_\unlhd L=\{v\mid v \unlhd w \text{ for some } w\in L\},$$
$$L_\unlhd=\{v\mid w \unlhd v \text{ for some }\ w\in L\}.$$
The \emph{freeness} operation, $L^\unlhd$ can defined for a language $L$, by
$$L^\unlhd \subseteq L \text{ and } \forall w\in L^\unlhd,
\forall v\in \Sigma^\star,\; v\lhd w\text{ implies }v\notin L^\unlhd.$$
The following proposition is
from~\cite{ang09:_languag_convex_with_respec_to}, except for the last
item.
\begin{propos}
Let $\unlhd$ be an arbitrary relation on $\Sigma^\star$. Then 
\begin{enumerate}
\item  A language is $\unlhd$-convex if and only if it is
  $\unrhd$-convex.
\item  A language is $\unlhd$-free if and only if it is $\unrhd$-free.
\item  Every $\unlhd$-closed language and every $\unrhd$-closed
  language is $\unlhd$-convex. 
\item  A language is $\unlhd$-closed if and only if its complement is
  $\unrhd$-closed.
 \item A language L is $\unlhd$-closed ($\unrhd$-closed) if and only
   if $L = _\unlhd L$ ($L = L_\unlhd$).
 \item A language L is $\unlhd$-free if and only   if $L = L^\unlhd$.
\end{enumerate}
\end{propos}

\noindent We consider $\unlhd$ to be:

\begin{itemize}
\item $\leq$: if $u,v,w\in \Sigma^\star$ and $w=uv$, then $u$ is
  \emph{prefix} of $w$, and we write $u\leq w$.
\item $\preceq$:  if $u,v,w\in \Sigma^\star$ and $w=uv$, then $v$ is
  \emph{suffix} of $w$, and we write $v\preceq w$ 
\item $\sqsubseteq$: if $u,v,w\in \Sigma^\star$ and $w=uxv$, then $x$ is
  \emph{factor} of $w$, and we write $x\sqsubseteq w$.  Note that a prefix or
  suffix of $w$ is also a factor of $w$. This relation is also  called \emph{infix}.
\item $\Subset$: if $w=w_0a_1w_1\cdots a_nw_n$, where $a_1,\ldots,a_n \in \Sigma$, and
$w_0,\ldots,w_n\in \Sigma^\star$, then $v = a_1\cdots a_n$ is a
\emph{subword} of $w$; and we write $v\Subset w$. Note that every
factor of $w$ is a subword  of $w$.
\end{itemize}

If a language is both prefix- and suffix-convex it is
\emph{bifix-convex}. In the same way are defined \emph{bifix-free} and
\emph{bifix-closed} languages. Ideals are languages directly related with closed languages. A
non-empty language $L\subseteq \Sigma^\star$ is a
\begin{itemize}
\item \emph{right ideal} if $L=L\Sigma^\star$ (also called \emph{ultimate
  definite} \cite{paz65:_ultim_defin_and_symmet_defin}); the
  complement is prefix converse-closed.
\item \emph{left ideal} if $L=\Sigma^\star L$ (also called \emph{reverse
  ultimate definite} \cite{paz65:_ultim_defin_and_symmet_defin}); the
  complement is suffix converse-closed.
\item \emph{two-sided ideal} if $L=\Sigma^\star L\Sigma^\star$ (also
  called \emph{central definite}); the complement is bifix converse-closed.
\item \emph{all-sided ideal} if $L=\Sigma^\star \shuffle L$; the complement
  is subword converse-closed; also studied by Haines ~\cite{haines69:_free_monoid_partial_order_by_embed} and
  Thierrin~\cite{thierrin72:_convex_languag}.
\end{itemize}

\begin{table}[htbp]
  \centering
\begin{tabular}{|c|c||c|c||c|c|}\hline
  \multicolumn{6}{|c|}{Free}\\\hline
  \multicolumn{1}{|c}{$\leq$}&$|\Sigma|$&\multicolumn{1}{c}{$\preceq$}&$|\Sigma|$&\multicolumn{1}{c}{$\sqsubseteq$}&$|\Sigma|$
\\\hline 
  $2^{m-1}+1$&3&$2^{m-1}+1$&$3$&$2^{m-2}+2$&$3$\\\hline
\multicolumn{2}{|c||}{$]m,2^{m-1}+1]$}&\multicolumn{2}{|c||}{$]m,2^{m-1}+1]$}&\multicolumn{2}{|c|}{$]m,2^{m-2}+2]$}\\\hline
  \multicolumn{6}{|c|}{Closed}\\\hline
  \multicolumn{1}{|c}{$\leq$}&$|\Sigma|$&\multicolumn{1}{c}{$\preceq$}&$|\Sigma|$
  &\multicolumn{1}{c}{$\sqsubseteq$}&$|\Sigma|$
\\\hline
  $2^m$&$3$&$2^{m-1}+1$&$4$&$2^{m-1}+1$&$4$\\\hline
\multicolumn{2}{|c||}{$]m,2^{m}]$}&\multicolumn{2}{|c||}{$[m,2^{m-1}+1]$}&\multicolumn{2}{|c|}{$]m,2^{m-1}+1]$}\\\hline
  \multicolumn{6}{|c|}{Ideal}\\\hline
  \multicolumn{1}{|c}{right}&$|\Sigma|$&\multicolumn{1}{c}{left}&$|\Sigma|$
  &\multicolumn{1}{c}{two-sided}&$|\Sigma|$
\\\hline
  $2^{m-1}$&$2$&$2^{m-1}+1$&$3$&$2^{m-2}+1$&$3$\\\hline
\end{tabular}
\caption{\small State complexity of determination of free, closed and ideal
  languages considering prefix, suffix and factor partial orders,
  respectively. For each free and closed of languages, the range of correspondent non-magic numbers appears on the second row.}
  \label{tab:scdeterminationconvex}
\end{table}

Some of the languages defined above are also characterized in terms of
properties of the finite automata that accept them. In particular:
prefix-closed languages are accepted by \nfas where all states are
final; suffix-closed languages are accepted by \nfas where all states
are initial; factor-closed languages are accepted by \nfas where all
states are initial and final; prefix-free languages are accepted by
non-exiting \nfas (i.e. there are no transitions from the final
states); suffix-free languages are accepted by non-returning \nfas
(i.e. there are no transitions to the initial state); and factor-free
languages are accepted by non-returning and non-exiting \nfas.

 The state complexity of the determination on some subregular
 languages (or for the kind of \nfas they are defined by) was
 recently studied by Bordihn \emph{et
   al.}~\cite{bordihn09:_deter_of_finit_autom_accep_subreg_languag},
 Jui-Yi Kao \emph{et al.}~\cite{kao09:_nfas_where_all_states_are}, and
 Jir\'askov\'a \emph{et
   al.}~\cite{jiraskova10:_compl_in_prefix_free_regul_languag}.
 Table~\ref{tab:scdeterminationconvex} presents some of the values for
 the languages considered above.  The existence of magic numbers for
 some subregular languages was studied by Holzer \emph{et
   al.}~\cite{holzer12:_magic_number_probl_for_subreg_languag_famil}. As
 can be seen in Table~\ref{tab:scdeterminationconvex}, $m$ is the only
 magic number for all free languages and for both prefix- and
 factor-closed languages (except if $m=1$, where $m$ is
 non-magic). Suffix-closed languages have no magic numbers.
\begin{table}[htbp]
  \centering
\begin{tabular}{|c|c|c|c|c|}\hline
\multicolumn{5}{|c|}{Prefix-free}\\\hline
&\multicolumn{1}{c}{sc}&$|\Sigma|$&\multicolumn{1}{c}{nsc}&$|\Sigma|$\\\hline
$L_1\cup L_2$&$mn-2$&$2$&$m+n$&$2$\\\hline
$L_1\cap L_2$&$mn-2(m+n-3)$& $2$  &$mn-(m+n)+2$&$2$\\\hline
$\comp{L}$&$m$&$1$&$2^{m-1}$&$3$\\\hline
 $(L_1- L_2)$&$mn-m-2n+4$ &$3$&$(m-1)2^{n-1}+1$&$4$
  \\\hline 
  $(L_1\oplus L_2)$&$mn-2$&$2$&&\\\hline
$L_1L_2$&$m+n-2$&$1$&$m+n-1$&$1$\\\hline
$L_1/L_2$&
\begin{tabular}{c}
$n-1$\\
$n-m+2$
\end{tabular}&
\begin{tabular}{c}
$2$ \\
$1$
\end{tabular}
&&\\\hline
$L^\star$&
\begin{tabular}{c}
$m$\\
$m-2$
\end{tabular}
&\begin{tabular}{c}
$2$ \\
$1$
\end{tabular}&
$m$&$1$\\\hline
$L^{R}$&$2^{m-2}+1$&$3$&$m$&$1$\\\hline
$\shift{L}$&$(2m-3)^{m-2}$&$6$&$2m^2-4m+3$&$2$\\\hline
\end{tabular}
  \caption{\small State complexity and nondeterministic state complexity of some operations on prefix-free languages}
  \label{tab:scnscprefixfree}
\end{table}

\paragraph{Free languages}
\label{sec:freelanguages}

Table~\ref{tab:scnscprefixfree} summarizes state complexity results of
individual operations on prefix-free
languages~\cite{han09:_nondet_state_compl_of_basic,
han09:_operat_state_compl_of_prefix,
jiraskova10:_compl_in_prefix_free_regul_languag,
brzozowski11:_quotien_compl_of_bifix_factor,
jiraskova14:_kleen_closur_regul_and_prefix_free_languag,
jiraskova14:_complement_pref_free,
jirasek14:_prefix_free_languag,
eom13:_state_compl_of_k_union}.
In the case of state complexity, the results are valid for Boolean operations  if
$m,n\geq 3$; for catenation if $m,n\geq 2$; for star if  $k=1$, then
$m\geq 3$, if $k=2$ then $m\not=3$, and else $m\geq 2$; and for reversal if
$m\geq 4$ and the tight bound cannot be reached if
$k=2$~\cite{jiraskova10:_compl_in_prefix_free_regul_languag}. The state complexty of right quotient is $1$, if  $k=1$ and $m=1$ or $m>n$, and if $k=2$ and $m=1$ or  $n=1$; furthermore, if $m=2$ then $sc(L_1/L_2)=n$~\cite{jirasek14:_prefix_free_languag}.

Note
that here the state complexity of the catenation and the star are much lower
than on general regular languages. Moreover, for the star, the only complexities attained are $m-2$, $m-1$, and $m$~\cite{jiraskova14:_kleen_closur_regul_and_prefix_free_languag}.

\begin{table}[htbp]
\centering
\begin{tabular}{|c||c|c|c|c|}\hline
\multicolumn{5}{|c|}{Suffix-free}\\\hline
&\multicolumn{1}{c}{sc}&$|\Sigma|$&\multicolumn{1}{c}{nsc}&$|\Sigma|$\\\hline
$L_1\cup L_2$&$mn-(m+n-2)$&$2$&$m+n-1$&$2$\\\hline
$L_1\cap L_2$&$mn-2(m+n -3)$&$2$&$mn-(m+n-2)$&$2$\\\hline
$\comp{L}$&$m$&$1$&
\begin{tabular}{c}
$2^{m-1}$\\
$\leq 2^{m-1}+2^{m-3}+1$\\
$\Theta(\sqrt{m})$
\end{tabular}
&
\begin{tabular}{c}
$3$\\
$2$\\
$1$
\end{tabular}
\\\hline
$L_1- L_2$&$mn-(m+2n-4)$&$4$&&\\\hline
$L_1\oplus L_2$&$mn-(m+n-2)$&$5$&&\\\hline
$L_1L_2$&$(m-1)2^{n-2}+1$&$4$&&\\\hline
$L^\star$&$2^{m-2}+1$&$4$&&\\\hline
$L^{R}$&$2^{m-2}+1$&$3$&&\\\hline
\end{tabular}
\caption{\small State complexity and nondeterministic state complexity of some operations on suffix-free languages}
  \label{tab:scncssuffixlanguages}
\end{table}
Table~\ref{tab:scncssuffixlanguages} summarizes the state complexity
of some regular operations on suffix-free languages.  Han and Salomaa
showed that all bounds, except for complementation, difference, and symmetric
difference, are tight
\cite{han09:_state_compl_of_basic_operat,han10:_nondet_state_compl_for_suffix}. Jir\'askov\'a
and Olej\'ar~\cite{jiraskova09:_state_compl_of_union_and} provided
binary witnesses for intersection and union. They also proved that for
all integer $\alpha$ between $1$ and the respective bound there are
languages $L_1$ and $L_2$ such that $(n)sc(L_1\circ L_2)=\alpha$, for
$\circ\in\{\cap,\cup\}$ (and witnesses ternary, except for $nsc(L_1\cap
L_2)$ for which the witnesses are over a four-symbol alphabet). The bounds for
difference and symmetric difference are from Brzozowski \emph{et al.}~\cite{brzozowski11:_quotien_compl_of_bifix_factor}. Jirásková \emph{et al.}~\cite{jiraskova14:_complement_pref_free} proved the results for complementation.

\begin{table}[htbp]
  \centering
\begin{tabular}{|c||p{4.5cm}|c|c|c|}\hline
\multicolumn{5}{|c|}{Free}\\\hline
&&\multicolumn{1}{|c}{$\leq\cup\preceq$}&\multicolumn{1}{|c}{$\sqsubseteq$}
&\multicolumn{1}{|c|}{$\Subset$}\\\hline
&\multicolumn{1}{|c|}{sc}&\multicolumn{3}{|c|}{$|\Sigma|$}\\\hline
$L_1\cup  L_2$&$mn-m-n$&$5$&$5$&$< m+n-3$\\\hline
$L_1\cap L_2$
&$mn-3m-3n+12$, $m,n\geq 4$&$3$&$3$&$m+n-7$\\\hline
$L_1-  L_2$&$mn-2m-3n+9$&$4$&$4$&$<  m+n-6$\\\hline
$L_1\oplus  L_2$&$mn-m-n$&$5$&$5$&$m+n-3$\\\hline
$L_1L_2$&$m+n-2$, $m,n>1$&1&1&1\\\hline
$L^\star$&$m-1$, $m>2$&2&2&2\\\hline
$L^{R}$&$2^{m-3}+2$, $m\geq 3$&2&2&$2^{m-3}-1$\\\hline
\end{tabular}
\caption{\small State complexity of basic operations on bifix-, factor-, and subword-free languages}
  \label{tab:scfixlanguages}
\end{table}

If a language is subword-free then it is factor-free, and if it is
factor-free then it is bifix-free.  Table~\ref{tab:scfixlanguages}
summarizes the state complexity of some regular operations on bifix-,
factor-, and subword-free
languages~\cite{brzozowski11:_quotien_compl_of_bifix_factor}.  The
tight upper bounds for the state complexity of these operations on the
three classes of languages coincide.

\begin{table}[htbp]
  \centering
\begin{tabular}{|c||c|c||c|c||c|c|c|}\hline
  \multicolumn{8}{|c|}{Closed}\\\hline
  &\multicolumn{1}{c}{$\leq$}&$|\Sigma|$&\multicolumn{1}{c}{$\preceq$}&
$|\Sigma|$&\multicolumn{1}{c}{$\sqsubseteq$,$\Subset$}&\multicolumn{1}{c}{$|\Sigma|_{\sqsubseteq}$}&\multicolumn{1}{c|}{$|\Sigma|_{\Subset}$}\\\hline
  $L_1\cup L_2$&$mn$&$2$&$mn$&$4$&$mn$&$2$&$2$\\\hline
  $L_1\cap  L_2$&$\scriptstyle{mn-m-n+2}$&$2$&$mn$&$2$&$\scriptstyle{mn-m-n+2}$&$2$&$2$\\\hline
  $L_1-L_2$&$mn-n+1$&$2$&$mn$&$4$&$mn-n+1$&$2$&$2$\\\hline
  $L_1\oplus
  L_2$&$mn$&$2$&$mn$&$2$&$mn$&$2$&$2$\\\hline
  $L_1L_2$&$\scriptstyle{m2^{n-2}+2^{n-2}}$&$3$&$\scriptstyle{mn-fn+f}$&$3$&$m+n-1$&$2$&$2$\\\hline
  $L^\star$&$2^{m-2}+1$&$3$&$m$&$2$&$2$&$2$&$2$\\\hline
  $L^R$&$2^{m-1}$&$2$&$2^{m-1}+1$&$3$&$2^{m-2}+1$&$3$&$2m$\\\hline\hline
  $_\unlhd  L$&$m$&$1$&$2^{m-1}$&$2$&$2^m-1$&$2$&\\\hline
$_\Subset  L$&&&&&\begin{tabular}{c}
$2^{m-2}+1$\\
$2^{\Omega(\frac{m}{3})}$
\end{tabular}&&
\begin{tabular}{c}
$m-2$\\
$2$
\end{tabular}
\\\hline
\end{tabular}
\caption{\small State complexity of some operations on prefix-, suffix-,
  factor-, and subword-closed languages. 
  The last two columns correspond to factor and subword, respectively.
  The last but one row contains the state complexity of the  closure of
  prefix, suffix, and factor respectively. The last row contains the state
 complexity of the subword closure, considering unbounded and binary alphabets, respectively.}
  \label{tab:scclosedlanguages}
\end{table}

\paragraph{Closed Languages and Ideals}
\label{sec:closedlanguagesideals}

Table~\ref{tab:scclosedlanguages} shows the state complexity 
of some basic operations on prefix-, suffix-, factor-, and
subword-closed
languages.  A
language is factor-closed if and only if it is subword-closed. So the
state-complexity results of operations are the same for those classes.
The state complexity of the closure on the respective partial orders
is also considered. Subword and converse subword closures were first
studied by 
Gruber \emph{et al.}~\cite{gruber07:_size_of_higman_haines_sets,gruber09:_more_size_of_higman_haines_sets}
and Okhotin~\cite{okhotin10:_state_compl_of_scatt_subst_and_super}.
Brzozowski \emph{et al.}~\cite{brzozowski10:_quotien_compl_of_closed_languag,brzozowski14:_quotien_compl_of_closed_languag} presented the tight upper bound, but using a growing alphabet. Karandikar and   Schoebelen~\cite{karandikar14:_state_compl_of_closur_and} shown that the exponential blown up is also required in the binary case. Given a regular
language $L$ with $sc(L)=m$, $nsc(_\Subset L)=nsc({L}_\Subset)=m$ and
these upper bounds are tight for witness binary languages.
Prefix, suffix, and factor closures (respectively, $_\leq L$, $_\preceq L$, and $_\sqsubseteq L$) were studied by Kao \emph{et
  al.}~\cite{kao09:_nfas_where_all_states_are}. 
If $L$ does not have $\emptyset$ as a quotient, Brzozowski \emph{et al.} shown that the state complexity of the suffix closure is $2^m -1$ (instead of $2^{m-1}$).
  
\begin{table}[htbp]
  \centering
\begin{tabular}{|c||c|c||c|c||c|c|c|}\hline
  \multicolumn{8}{|c|}{Ideal}\\\hline
  &\multicolumn{1}{c}{right}&$|\Sigma|$&
\multicolumn{1}{c}{left}&$|\Sigma|$&
\multicolumn{1}{c}{-sided}&\multicolumn{1}{c}{$|\Sigma|_{\text{two}}$}&$|\Sigma|_{\text{all}}$\\\hline
  $L_1\cup
  L_2$&$\scriptstyle{mn-m-n+2}$&$2$&$mn$&$4$&$\scriptstyle{mn-m-n+2}$&$2$&$2$\\\hline
  $L_1\cap
  L_2$&$mn$&$2$&$mn$&$2$&$mn$&$2$&$2$\\\hline
  $L_1-L_2$&$\scriptstyle{mn-m+1}$&$2$&$mn$&$4$&$\scriptstyle{mn-m+1}$&$2$&$2$\\\hline
  $L_1\oplus
  L_2$&$mn$&$2$&$mn$&$2$&$mn$&$2$&$2$\\\hline
  $L_1L_2$&$\scriptstyle{m+2^{n-2}}$&$1$&$\scriptstyle{m+n-1}$&$1$&$\scriptstyle{m+n-1}$&$1$&$3$\\\hline
  $L^\star$&$m+1$&$2$&$m+1$&$2$&$m+1$&$2$&$2$\\\cline{2-8}
&\multicolumn{7}{|l|}{If $\varepsilon\in L$, then $L=\Sigma^\star$ and
  $sc(L^\star)=1$.}\\\hline
  $L^R$&$2^{m-1}$&$2$&$2^{m-1}+1$&$3$&$2^{m-2}+1$&$3$&$\scriptstyle{2m-4}$\\\hline
 \end{tabular}
  \caption{\small State complexity of basic operations on ideals. The last
    two columns correspond to two-sided  and all-sided ideals, respectively.}
  \label{tab:scideals}
\end{table}

If $L$ is a right (respectively, left, two-sided, all-sided) ideal, any
language $G \subseteq \Sigma^\star$ such that $L=G\Sigma^\star$
(respectively, $L=\Sigma^\star G$, $L=\Sigma^\star
G\Sigma^\star$,$L=\Sigma^\star\shuffle	G$) is a \emph{generator} of $L$.  Brzozowski and
Jir\'askov\'a~\cite{brzozowski10:_quotien_compl_of_ideal_languag} studied
state complexity on ideals. Table~\ref{tab:scideals} presents the
state complexity of basic operations on ideals.  As stated before
closed languages and ideals are related. In particular, the state
complexity of basic operations on two-sided and all-sided ideals
coincide. Brzozowski~\cite{brzozowski10:_compl_in_convex_languag}
observed that for the four types of convex languages (prefix, suffix,
factor and subword) the state complexity of the Boolean operations is
$mn$.

\paragraph{Unary convex languages}
In the case of unary languages, prefix, suffix, factor, and subword partial orders
coincide. Table~\ref{tab:scunaryfreeclosedideals} summarizes the state
complexity of basic operations on unary free, unary closed, unary
ideals and unary convex languages.

\begin{table}[htbp]
  \centering
\begin{tabular}{|c||c|c|c|c|}\hline
  \multicolumn{5}{|c|}{Unary}\\\hline
  &\multicolumn{1}{c|}{Free}&\multicolumn{1}{c|}{Closed}&\multicolumn{1}{c|}{Ideal}&\multicolumn{1}{c|}{Convex}\\\hline
  $L_1\cup L_2$&$\max\{m,n\}$&$\max\{m,n\}$&$\min\{m,n\}$&$\max\{m,n\}$\\\hline
  $L_1\cap L_2$&$m=n$&$\min\{m,n\}$&$\max\{m,n\}$&$\max\{m,n\}$\\\hline
  $L_1-L_2$&$m$&$m$&$n$&$\max\{m,n\}$\\\hline
  $L_1\oplus  L_2$&$\max\{m,n\}$&$\max\{m,n\}$&$\max\{m,n\}$&$\max\{m,n\}$\\\hline
  $L_1L_2$&$m+n-2$&$m+n-2$&$m+n-1$&$m+n-1$\\\hline
  $L^\star$&$m-2$&$2$&$m-1$&$n^2-7n+13$\\\hline
  $L^R$&$m$&$m$&$m$&$m$\\\hline
\end{tabular}
  \caption{\small State complexity of basic operations on unary convex languages}
  \label{tab:scunaryfreeclosedideals}
\end{table}

\begin{table}[htbp]
  \centering
\begin{tabular}{|c||c|c||c|c|}\hline
&\multicolumn{2}{|c||}{Regular}&\multicolumn{2}{c|}{Ideal}\\\hline
&\multicolumn{1}{c}{sc}&$|\Sigma|$&\multicolumn{1}{c}{sc}&$|\Sigma|$\\\hline
$L^\leq$&$m+1$&$2$&$m+1$&$2$\\\hline
$L^\preceq$&$(m-1)2^{m-2}+2$, $m\geq 4$&$4$&$\frac{n(n-1)}{2}+2$&$1$\\\hline
$L^\sqsubseteq$&$(m-2)2^{m-3}+3$, $m\geq 4$&$3$&$n+1$&$1$\\\hline
$L^b$& $(m-2)2^{m-2}+3$, $m\geq4$&$4$&&\\\hline
\end{tabular}
  \caption{\small State complexity  of prefix, suffix, factor and bifix
    operations on  regular languages and on ideals (right, left and
    two sided, respectively).}
  \label{tab:scfreeness}
\end{table}

\paragraph{Freeness Operations} 
Here we analyse the state complexity of freeness
operations for prefix, suffix, bifix and factor orders that were
studied by Pribavkina and
Rodaro~\cite{pribavkina10:_state_compl_of_prefix_suffix}.
Given a regular language $L$, the $\unlhd$-free language
$L^\unlhd$ for $\unlhd\in\{\leq,\preceq,\sqsubseteq\}$, is  respectively\footnote{In~\cite{pribavkina10:_state_compl_of_prefix_suffix}
the superscripts for prefix, suffix and factor operations were respectively $p$,
$s$ and~$\iota$.}:
\begin{itemize}
\item prefix: $L^\leq=L-L\Sigma^+$
\item suffix: $L^\preceq=L-\Sigma^+L$
\item factor: $L^\sqsubseteq=L-(\Sigma^+ L \Sigma^\star
  \cup \Sigma^\star L \Sigma^+ )$
\end{itemize}
\noindent The bifix operation is defined by $L^b=L^\leq \cap
L^\preceq$.  If $L$ is an ideal, prefix, suffix and factor operations
were studied by Brzozowski and
Jir\'askov\'a~\cite{brzozowski10:_quotien_compl_of_ideal_languag}. In this
case, the resulting languages are minimal generators for left, right
and two sided ideals, respectively. Table~\ref{tab:scfreeness}
presents the state complexity of prefix, suffix, factor and bifix
operations on regular languages (and correspondent ideals). The state
complexity of this operations is much lower in the case of right and
two-sided ideals than for general regular languages.

\subsubsection{Star-free Languages}
\label{sec:starfree}
\begin{figure}[hbt]
     \centering

    \includegraphics[width=8cm]{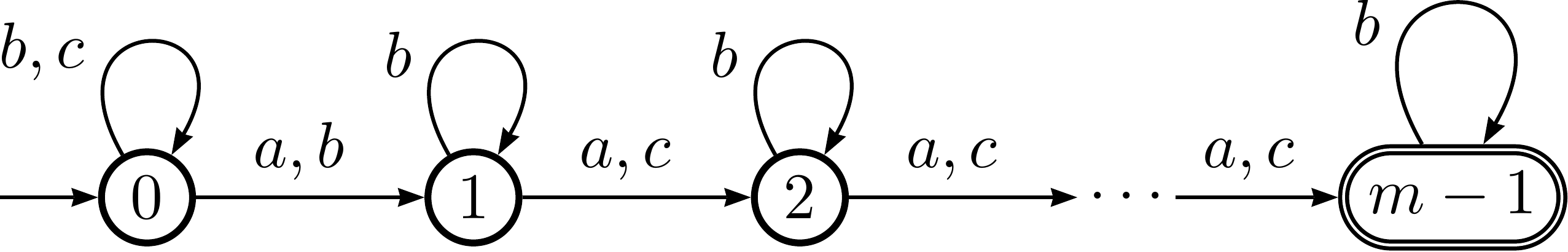}
		\ignore{\SmallPicture{\VCDraw{
              \begin{VCPicture}{(-2,-1)(14,2)}
                \State[0]{(0,0)}{0}\Initial{0}
                \State[1]{(3,0)}{1}
                \State[2]{(6,0)}{2}
                \SetStateLineStyle{none}
                \State[\cdots]{(9,0)}{3}
                \SetStateLineStyle{solid}
                \FinalStateVar[m-1]{(12,0)}{m1}
                \LoopN{0}{b,c}\EdgeL{0}{1}{a,b}
                \LoopN{1}{b}\EdgeL{1}{2}{a,c}
                \LoopN{2}{b}\EdgeL{2}{3}{a,c}
                \EdgeL{3}{m1}{a,c}
                \LoopN{m1}{b}
              \end{VCPicture}
             }
           }  }   

  \caption{\small{}Minimal $m$-state \nfas with equivalent minimal $2^m$-state \dfa for star-free languages}
     \label{fig:starfreedet}
   \end{figure}

Star-free languages are the smallest class containing the finite
languages and closed under Boolean operations and catenation. This
class of languages correspond exactly to the regular languages of star
height $0$. The minimal \dfas of star-free languages are
\emph{permutation-free} (i.e. no word performs a non-trivial
permutation of a subset of its states).  Bordhin \emph{et
  al.}~\cite{bordihn09:_deter_of_finit_autom_accep_subreg_languag}
showed that the state complexity of the determination of a star-free
language $L$ is $2^{nsc(L)}$. Figure~\ref{fig:starfreedet} presents a
family of ternary \nfas for which the bound is tight. Holzer \emph{et
  al.}~\cite{holzer12:_magic_number_probl_for_subreg_languag_famil}
showed that star-free languages have no magic numbers.

   Brzozowski and
   Liu~\cite{brzozowski11:_quotien_compl_of_star_free_languag} studied
   the state complexity of the basic regular operations on star-free
   languages, and their results are summarized in
   Table~\ref{tab:starfree}. The bounds obtained for general regular
   languages are reached except in the catenation for $n=2$, the
   reversal, and operations on unary languages. Holzer \emph{et
     al.}~\cite{holzer11:_nondet_state_compl_of_star_free_languag,holzer12:_nondet_state_compl_of_star_free_languag}
   studied the same languages for the operational nondeterministic
   state complexity. The bounds  coincide with the ones for general
   regular languages and are tight for binary languages. The witness languages
   for union and catenation are $a^{m-1}(ba^{m-1})^\star$ and
   $b^{n-1}(ab^{n-1})^\star$. For intersection, witnesses are  $b^\star(ab^\star)^{m-1}$ and
   $a^\star(ba^\star)^{n-1}$. The first witness for union is also a
   witness for the star operation. The language family  presented in
   Figure~\ref{fig:jirasreveralnfa} is star-free and thus  a witness for the reversal operation.
  On unary star-free languages, the upper bounds for  operational nondeterministic
   state-complexity coincide with general case, except for the
   complementation. Holzer \emph{et
     al.}~\cite{holzer12:_nondet_state_compl_of_star_free_languag}
   showed that for reversal and star the bounds are tight.  For union, the presented lower bound
misses the upper bound by one state. For intersection, the presented
bound is tight in the order of magnitude ($\Theta(mn)$) and the bound for
complementation is $\Theta(n^2)$. The  lower bound  for catenation  misses the upper
bound for unary general languages by one state.

\begin{table}[htbp]
  \centering
\begin{tabular}{|l||c|c||c|}\hline
\multicolumn{4}{|c|}{Star-free}\\\hline
&\multicolumn{1}{c}{sc}&$|\Sigma|$&\multicolumn{1}{c|}{Unary}\\\hline\hline
$L_1\circ L_2$&$mn$&2&$\max\{m,n\}$\\\hline
\multirow{2}{*}{$L_1L_2$}&$(m-1)2^n+2^{n-1}$, if $n\geq 3$&$4$&\multirow{2}{*}{$m+n-1$}\\\cline{2-3}
&$[3m-2,3m-1]$, if $n=2$
&$3$&\\\hline
\multirow{3}{*}{$L^\star$}& $2$, if $m=1$&$1$& $2$, if $m=1$\\\cline{2-4}
&$2^{m-1}+2^{m-2}$, if $m\geq 2$&$4$&$m$, if $m\in[2,5]$\\\cline{2-4}
&&&$m^2-7m+13$, if $m>5$
\\\hline
$L^R$&$2^m-1$&$m-1$&$m$\\\hline
\end{tabular}
\caption{\small State complexity  of basic regular operations on star-free
  regular and unary languages, where $\circ\in
  \{\cup,\cap,\setminus,\oplus\}$. For non-unary star-free languages
  and $n=2$, $m\geq 2$.
  For non-unary star-free languages if $m\in [1,2]$, the bound for
  reversal is tight  for $|\Sigma|\geq m$, and if $m\geq 3$, for $|\Sigma|\geq m-1$.}
  \label{tab:starfree}
\end{table}

\subsection{Some More Results}
We briefly cite some more work on operational state complexity.
C\^ampeanu and Ho~\cite{campeanu04:_maxim_state_compl_for_finit_languag}
and Brzozowski and
Konstantinidis~\cite{brzozowski09:_state_compl_hierar_of_unifor}
considered uniform finite languages.  Krieger \emph{et al.} studied
decimations of
languages~\cite{krieger09:_decim_of_languag_and_state_compl}.
C\^ampeanu and
Konstantinidis~\cite{campeanu08:_state_compl_of_subwor_closur}
analysed a subword closure operation.  Union-free languages were
considered by Jir\'askov\'a and
Masopust~\cite{jiraskova10:_compl_in_union_free_regul_languag,jiraskova11:_compl_in_union_free_regul_languag}.
The same authors studied the state complexity of projected languages~\cite{jiraskova11:_state_compl_of_projec_languag}.
The chop (or \emph{fusion}) of two words is their catenation where the
touching symbols are merged if equal, or is undefined otherwise. The
chop operation and its iterated variants (star and plus) where studied
by Holzer \emph{et al.}~\cite{holzer11:_chop_of_languag,
holzer11:_chop_operat_and_expres,holzer12:_state_compl_of_chop_operat}. The
(nondeterministic) state complexity results are similar to  the ones
for catenation, star and plus, with the exception of chop-star where
the complexities also depend on the alphabet size. This  comes as a
surprise as chop based regular expressions are known to be
exponentially more succinct than classical catenation based ones. Bassino \emph{et al.}~\cite{bassino10:_compl_of_operat_cofin_languag}
provided upper bounds of the state complexity of basic operations on
cofinite languages as a function of the size the of complementary finite
language (taken as the summation of the lengths of all its words).  The
average state complexity on finite languages is addressed in two
works.  Gruber and
Holzer~\cite{gruber07:_averag_state_and_trans_compl} analysed the
average state complexity of \dfas and \nfas based on a uniform
distribution over finite languages whose longest word is of length at
most $n$.  Based on the size of finite languages as the summation of
the lengths of all its words and a correspondent uniform distribution,
Bassino \emph{et al.}~\cite{bassino10:_averag_state_compl_of_ration}
establish that the average state complexities of the basic regular
operations are asymptotically linear.

\section{State Complexity of Combined Operations}
\label{sec:scco}
The number of standard individual operations on regular languages is clearly
limited and almost all of their state complexities have been already
obtained. However, in many practical cases, not only these
individual operations but also their combinations are used,
for example, the operations expressed by the regular expressions in
the programming language Perl. These combinations are called
combined operations.

In 2011, Salomaa \emph{et al.} \cite{SaSaYu11} proved that it cannot exist an algorithm such that, for a given composition of basic regularity preserving operations, computes the state complexity of the corresponding composed operation. The undecidability result holds already for arbitrary compositions of intersection and marked concatenation and the proof relies on a reduction from Hilbert's Tenth Problem. Although the composition of
state complexities of individual component operations of a combined operation
would give an upper bound for the state complexity of the
combined operation, the upper bound is usually too high to be
meaningful~\cite{LiMaSaYu08,salomaa07a:_state_compl_of_combin_operat,Yu06}. For example, for two regular
languages $L_1$ and $L_2$ accepted by $m$-state and an $n$-state
\dfa, respectively, the exact state
complexity of $(L_1\cup L_2)^*$ is actually
$2^{m+n-1}-2^{m-1}-2^{n-1}+1$, while the composition of their
individual state complexities is $2^{mn-1}+2^{mn-2}$. Clearly,
$O(2^{m+n})$ and $O(2^{mn})$ are totally different.

Since the number of combined operations is unlimited and the state
complexities of many of them are very difficult to
compute, it would be good if we have a general estimation
method that generates close upper bounds of the state complexities
of combined operations which are good enough
to use in practice. Such an estimation method has been proposed by Ésik \textit{et al.}~\cite{EsGaLiYu09}, and Salomaa and Yu~\cite{SaYu07}. A further concept in this direction,
approximation of state complexity has been introduced Gao and Yu~\cite{GaYu12}.

In the following, we will survey both the results of state
complexities of combined operations and the results of estimations
and approximations of state complexities of combined operations.

\subsection{State Complexity of Combined Operations on Regular Languages}

The state complexities of a number of basic combined operations on regular languages have been
studied. Most of these combined operations are composed of two basic individual operations. The results are shown in Table~\ref{tab:sc-some-combined-regular}.

In 1996, Birget~\cite{birget96:_state_compl_of_oever_sigma} obtained the the state complexity of $\overline{\Sigma^\star \overline{L}}$, where $L$ is a regular language. This combination of complementation, catenation and star is the first combined operation composed of different individual operations whose state complexity was established. In 2007, Salomaa \emph{et al.}~\cite{salomaa07a:_state_compl_of_combin_operat} pointed out that the mathematical composition of state complexities of individual component operations of a combined operation is usually much higher than the state complexity of the combined operation. This is because the result of a component operation of the combined operation may not be among the worst-cases of
the succeeding component operation. They established the state complexity of $(L_1 \cup L_2)^*$ and indicated that the state complexity of $(L_1 \cap L_2)^*$ should be at least reasonably close to the mathematical composition of state complexities of intersection and star. Later, Jir\'askov\'a and Okhotin~\cite{jiraskova11:_state_compl_of_star_of} proved that the state complexity of $(L_1 \cap L_2)^*$ is exactly the same as the mathematical composition of state complexities of intersection and star.

Gao \emph{et al.}~\cite{GaSaYu08},  in 2008, established the state complexities of $(L_1 L_2)^*$ and $(L_1^R)^*$, where $L_1$ and $L_2$ are regular languages. The state complexity of $(L_1 L_2)^*$ is $2^{m +n -1}-2^{m -1}-2^{n -1}+1$ which is lower than the mathematical composition of the state complexity of catenation and star. Interestingly, the state complexity of $(L_1^R)^*$ is the same as that of $L_1^R$ which is $2^m$. The worst-case example over a three-letter alphabet for $L_1^R$ \cite{yu94:_state_compl_of_some_basic} also works for $(L_1^R)^*$.

\begin{table}[htbp]
\begin{tabular}{|l||c|c|}
\hline
\multicolumn{3}{|c|}{Regular}\\
\hline
&\multicolumn{1}{c}{sc}&\multicolumn{1}{c|}{$|\Sigma|$}\\
\hline
\hline
$\overline{\Sigma^\star \overline{L_1}}$ & $2^{m-1}$ (\cite{birget96:_state_compl_of_oever_sigma}) & 2\\
\hline
$\overline{L_1^*}^*$ & $2^{\theta(m\log m)}$ (\cite{jiraskova12:_state_compl_of_star_compl_star}) & 7\\
\hline
$(L_1 \cup L_2)^*$ & $2^{m +n -1}-2^{m -1}-2^{n -1}+1$ (\cite{jiraskova11:_state_compl_of_star_of,salomaa07a:_state_compl_of_combin_operat}) & 2\\
\hline
$(L_1 \cap L_2)^*$ & $2^{mn -1}+2^{mn -2}$ (\cite{jiraskova11:_state_compl_of_star_of}) & 6\\
\hline
$(L_1 L_2)^*$   &  $2^{m+n-1}+2^{m+n-4}-2^{m-1}-2^{n-1}+m+1$ (\cite{GaSaYu08}) & 4\\\hline
$(L_1^R)^*=(L_1^*)^R$   &  $2^m$  (\cite{GaSaYu08}) & 3\\
\hline
$(L_1\cup L_2)^R$ & $2^{m+n}-2^m-2^n+2$ (\cite{LiMaSaYu08}) & 3\\
    \hline
$(L_1\cap L_2)^R$ & $2^{m+n}-2^m-2^n+2$ (\cite{LiMaSaYu08})  & 3\\
    \hline
$(L_1L_2)^R$ & $3\cdot 2^{m+n-2}-2^n+1$ (\cite{CuGaKaYu12,LiMaSaYu08}) & 4\\
    \hline
    $L_1^*L_2$   &  $5 \cdot 2^{m+n-3} - 2^{m-1} - 2^{n} +1$ (\cite{CuGaKaYu12})   & 4\\
\hline
$L_1L_2^*$   &  $(3m-1)2^{n-2}$ (\cite{CGKY12-cat-sr}) & 3\\
\hline
$L_1^RL_2$   &  $3\cdot 2^{m+n-2}$ (\cite{CuGaKaYu12}) & 4\\
\hline
$L_1L_2^R$   &  $m 2^{n}-2^{n-1}-m+1$ (\cite{CGKY12-cat-sr}) & 3\\
\hline
$L_1(L_2\cup L_3)$   &  $(m-1)(2^{n+p}-2^{n}-2^{p}+2)+2^{n+p-2}$ (\cite{CGKY11-cat-ui}) & 4\\
\hline
$L_1(L_2\cap L_3)$   &  $m 2^{np}-2^{np-1}$ (\cite{CGKY11-cat-ui}) & 4\\
\hline
$L_1^*\cup L_2$   &  $3\cdot 2^{m-2}\cdot n-n+1$ (\cite{GaYu10}) & 3\\
\hline
$L_1^*\cap L_2$   &  $3\cdot 2^{m-2}\cdot n-n+1$ (\cite{GaYu10})  & 3\\
\hline
$L_1^R\cup L_2$   &  $2^{m}\cdot n-n+1$ (\cite{GaYu10}) & 4\\
\hline
$L_1^R\cap L_2$   &  $2^{m}\cdot n-n+1$ (\cite{GaYu10}) & 4\\
\hline
$(L_1 \cup L_2)L_3$ & $mn2^p - (m+n-1)2^{p-1}$ (\cite{CuGaKaYu12}) & 4\\
\hline
$(L_1 \cap L_2) L_3$ & $mn2^{p}-2^{p-1}$ (\cite{CuGaKaYu12})
 & 4\\
\hline
$L_1L_2 \cup L_3$ & $(m2^{n}-2^{n-1})p$ (\cite{CuGaKaYu12}) & 4\\
\hline
$L_1L_2 \cap L_3$ & $(m2^n-2^{n-1})p$ (\cite{CuGaKaYu12}) & 3\\
\hline
$ L_1 L_2 L_3$ & $m2^{n+p}-2^{n+p-1}-(m-1)2^{n+p-2}$ & 5\\
    & $-2^{n+p-3}-(m-1)(2^p-1)$ (\cite{EsGaLiYu09}) & \\
\hline
\end{tabular}
  \centering
  \caption{\small{State complexities of some basic combined operations on regular languages}}\label{tab:sc-some-combined-regular}
\end{table}

In 2008, Liu \emph{et al.}~\cite{LiMaSaYu08} studied the state complexities of $(L_1\cup L_2)^R$, $(L_1\cap L_2)^R$, and $(L_1L_2)^R$, where $L_1$ and $L_2$ are regular languages. The tight bounds for $(L_1\cup L_2)^R$ was proved and the state complexity of $(L_1\cap L_2)^R$ is the same as that of $(L_1\cup L_2)^R$ because of De Morgan's laws and $\overline{L^R}=\overline{L}^R$. They also gave an upper bound for the last combined operation which was proved to be tight, in 2012, by Cui \emph{et al.}~\cite{CuGaKaYu12}.

Cui \emph{et al.}~\cite{CGKY11-cat-ui} established the state complexities of $L_1(L_2\cup L_3)$ and $L_1(L_2\cap L_3)$ in 2011. The state complexity of $L_1(L_2\cup L_3)$ is lower than the mathematical composition of the state complexities of union and catenation, whereas the state complexity of $L_1(L_2\cap L_3)$ is the same as the corresponding composition.

In 2012, Jir{\'a}skov{\'a} and Shallit~\cite{jiraskova12:_state_compl_of_star_compl_star} proved the state complexity of the combined operation $\overline{L_1^*}^*$ to be $2^{\theta(m\log m)}$, where $L_1$ is a regular language accepted by an $m$-state \dfa. A seven-letter alphabet was used in the proof for the lower bound.

Gao \emph{et al.} presented the state complexities of four combined operations: $L_1^*\cup L_2$, $L_1^*\cap L_2$, $L_1^R\cup L_2$, and $L_1^R\cap L_2$, where $L_1$ and $L_2$ are regular languages accepted by $m$ and $n$-state \dfas, respectively. The state complexities of the four combined operations are all $n-1$ less than the mathematical composition of the state complexities of their component operations. Although gaps are the same, the reasons causing them are different. For $L_1^*\cup L_2$ and $L_1^*\cap L_2$, the gap $n-1$ exists because there are $n-1$ unreachable states in the constructions of resulting \dfas. For $L_1^R\cup L_2$ and $L_1^R\cap L_2$, it is because $n$ states are equivalent and can be merged into one in the constructions.

Cui \emph{et al.}~\cite{CGKY12-cat-sr,CuGaKaYu12} gave the state complexities of a number of combined operations including: $L_1^*L_2$, $L_1L_2^*$, $L_1^RL_2$, $L_1L_2^R$, $(L_1 \cup L_2)L_3$, $(L_1 \cap L_2)L_3$, $L_1L_2 \cup L_3$, and $L_1L_2 \cap L_3$. The state complexities of the first five combined operations are less than the corresponding mathematical compositions and the state complexities of the others are the same as the compositions. The state complexity of $L_1L_2^R$ is equal to that of catenation combined with antimorphic involution $(L_1 \theta(L_2))$ in biology \cite{CGKY12-cat-sr}. Up to now, the state complexities of all the combined operations composed of two basic individual operations have been obtained. These results will serve as the basis of the research on the state complexities of combined operations with more complex structures in the future.

Besides these basic combined operations, a few combined operations on $k$ operand regular languages have also been investigated, e.g. $(\bigcup\limits_{i=1}^{k} L_i)^*$, $k\ge 2$. These results are summarized in Table~\ref{tab:sc-classes-of-combined-regular}. The state complexity of $L_1\cap L_2\cap \ldots \cap L_k$, $k\ge 2$ was shown to be $n_1n_2\cdots n_k$ by Birget \cite{birget91:_inter_of_regul_languag_and_state_compl}, and Yu  and Zhuang~\cite{yu91:_state_compl_of_inter_of_regul_languag} in 1991, where $L_i$ is a regular language accepted by an $n_i$-state \dfa, $1\le i\le k$. \'Esik \emph{et al.}~\cite{EsGaLiYu09} later extended the result to combined Boolean operations. A combined Boolean operation $f(L_1,L_2,\ldots,L_k)$ is a function which can be constructed from the projection functions and the binary union, intersection and the complementation operations by function composition, e.g. $\overline{L_1}\cup L_2\cap L_2\cap \ldots \cap L_k$. Its state complexity was proved to be also $n_1n_2\cdots n_k$.\ \'Esik \emph{et al.}~\cite{EsGaLiYu09} presented the state complexities of $L_1L_2L_3$ and $L_1L_2L_3L_4$ in the same paper. The worst-case examples for the two combined operations are modifications of the worst-case examples proposed by Yu \emph{et al.}~\cite{yu94:_state_compl_of_some_basic} for catenation. On the basis of these results, Gao~\cite{Ga10} established the state complexity of $L_1 L_2 \ldots L_k$, which formula is too complex to figure here.

In 2012, Gao \emph{et al.}~\cite{GaKaYu12-union-and-intersection-of-square-and-reversal} gave the state complexities of a series of combined operations composed of arbitrarily many individual operations, including: $(\bigcup\limits_{i=1}^{k} L_i)^*$, $(\bigcup\limits_{i=1}^{k} L_i)^2$, $\bigcup\limits_{i=1}^{k} L_i^*$, $\bigcap\limits_{i=1}^{k} L_i^*$, $\bigcup\limits_{i=1}^{k} L_i^2$, $\bigcap\limits_{i=1}^{k} L_i^2$, $\bigcup\limits_{i=1}^{k} L_i^R$, and $\bigcap\limits_{i=1}^{k} L_i^R$. Tight bounds were established for all these combined operations.

In Table~\ref{tab:sc-classes-of-combined-regular}, we can see that all the results on the state complexities of combined operations on $k$ operand languages were proved with increasing alphabets. Clearly, it is comparatively easier to design worst-case examples with increasing alphabets than fixed ones. However, the most crucial reason is that it is impossible to design a worst-case example for a combined operation on
arbitrary $k$ operand languages which are over a fixed alphabet and accepted by arbitrary $n_1$, $n_2$, $\ldots$, $n_k$-state \dfas, respectively. This is because there exist only a limited number of different \dfas
with a fixed number of states if the alphabet is fixed. Therefore, when $k$ is large enough and $n_i$ is an arbitrary positive integer, $1\le i\le k$,
some of the \dfas may have the same number of states and some of them may be indeed
the same according to pigeonhole principle \cite{GaKaYu12-union-and-intersection-of-square-and-reversal}. Thus, the research on the state
complexities of combined operations on $k$ operand languages uses increasing alphabets
in general.

\begin{table}[htbp]
\begin{tabular}{|l||c|c|}
\hline
\multicolumn{3}{|c|}{Regular}\\
\hline
&\multicolumn{1}{c}{sc}&\multicolumn{1}{c|}{$|\Sigma|$}\\
\hline
\hline
$(\bigcup\limits_{i=1}^{k} L_i)^*$ & $\prod\limits_{i=1}^{k}(2^{n_i-1}-1)+2^{\sum\limits_{j=1}^{k}n_j-k}$ (\cite{GaKa12}) & $2k+1$\\
\hline
$(\bigcup\limits_{i=1}^{k} L_i)^2$ & $\prod\limits_{h=1}^{k}(n_h-1)[\prod\limits_{i=1}^{k}(2^{n_i}-1)+1]$ & $2k+1$\\
  & $+[\prod\limits_{j=1}^{k}n_j -\prod\limits_{l=1}^{k}(n_l-1)]2^{\sum\limits_{m=1}^{k}n_m-k}$ (\cite{GaKa12}) & \\
\hline
$\bigcup\limits_{i=1}^{k} L_i^*$   &  $(\frac{3}{4}) ^k 2^{g}-\sum\limits_{i=1}^{k}[\prod\limits_{j=1}^{i-1}(\frac{3}{4}2^{n_j}-1)\prod\limits_{t=i+1}^{k}(\frac{3}{4}2^{n_t})]
+1$ (\cite{GaKaYu12-union-and-intersection-of-star}) & $2k$\\
\hline
$\bigcap\limits_{i=1}^{k} L_i^*$   &  $(\frac{3}{4}) ^k 2^{g}-\sum\limits_{i=1}^{k}[\prod\limits_{j=1}^{i-1}(\frac{3}{4}2^{n_j}-1)\prod\limits_{t=i+1}^{k}(\frac{3}{4}2^{n_t})]
+1$ (\cite{GaKaYu12-union-and-intersection-of-star}) & $2k$\\
\hline
$\bigcup\limits_{i=1}^{k} L_i^2$   &  $\prod\limits_{i=1}^{k}(n_i 2^{n_i}-2^{n_i-1})$ (\cite{GaKaYu12-union-and-intersection-of-square-and-reversal}) & $2k$\\
\hline
$\bigcap\limits_{i=1}^{k} L_i^2$   &  $\prod\limits_{i=1}^{k}(n_i 2^{n_i}-2^{n_i-1})$ (\cite{GaKaYu12-union-and-intersection-of-square-and-reversal}) & $2k$\\
\hline
$\bigcup\limits_{i=1}^{k} L_i^R$   &  $\prod\limits_{i=1}^{k} (2^{n_i} - 1) + 1$ (\cite{GaKaYu12-union-and-intersection-of-square-and-reversal}) & $3k$\\
\hline
$\bigcap\limits_{i=1}^{k} L_i^R$   &  $\prod\limits_{i=1}^{k} (2^{n_i} - 1) + 1$ (\cite{GaKaYu12-union-and-intersection-of-square-and-reversal}) & $3k$\\
    \hline
A Boolean
 & $n_1n_2\cdots n_k$ (\cite{birget91:_inter_of_regul_languag_and_state_compl,EsGaLiYu09,yu91:_state_compl_of_inter_of_regul_languag}) & $2k$\\
operation & & \\
$f(L_1,\ldots,L_k)$ & & \\
    \hline
    $L_1L_2\cdots L_k$   &  see details in \cite{EsGaLiYu09,Ga10,GaYu09}    & $2k-1$\\
\hline
\end{tabular}
  \centering
  \caption{\small{State complexities of some combined operations on $k$ regular languages, $k\ge 2$}}\label{tab:sc-classes-of-combined-regular}
\end{table}

\subsection{State Complexity of Combined Operations on Prefix-free Regular Languages}
Since the research history of combined operations is much shorter than that of individual operations, there remains a lot of work to be done on state complexity of combined operations for subregular language classes. The state complexities of several combined operations on prefix-free regular languages were obtained by Han \emph{et al.}~\cite{HaSaYu10}, in 2010. These results are shown in Table~\ref{tab:sc-combined-prefix-free}.

\begin{table}[htbp]
\begin{tabular}{|l||c|c|}
\hline
\multicolumn{3}{|c|}{Prefix-Free Regular}\\
\hline
&\multicolumn{1}{c}{sc}&\multicolumn{1}{c|}{$|\Sigma|$}\\
\hline
\hline
$(L_1 \cup L_2)^*$ & $5\cdot 2^{m +n -6}$ (\cite{HaSaYu10}) & 4\\
\hline
$(L_1 \cap L_2)^*$ & $mn-2(m+n)+6$ (\cite{HaSaYu10}) & 4\\
\hline
$(L_1 L_2)^*$   &  $m+n-2$ (\cite{HaSaYu10}) & 2\\
\hline
$(L_1^R)^*=(L_1^*)^R$   &  $2^{m-2}+1$ (\cite{HaSaYu10}) & 3\\
\hline
\end{tabular}
  \centering
  \caption{\small{State complexities of some combined operations on prefix-free regular languages}}\label{tab:sc-combined-prefix-free}
\end{table}

\subsection{Estimation and Approximation of State Complexity of
Combined Operations}
\label{sec:estapp}
We can summarize at least two problems concerning the state complexities for combined operations.
 First, the state complexities of combined operations composed of large numbers of individual operations are extremely difficult to compute. Second, a large proportion of results that have been obtained are pretty complex and impossible to comprehend~\cite{GaYu09}.
 For example, Ésik et al.~\cite{EsGaLiYu09} shown that the state complexity of the catenation for four
regular languages with state complexities $m, n, p, q$, respectively, is
$$9(2m-1)2^{n+p+q-5}-3(m-1)2^{p+q-2}-(2m-1)2^{n+q-2}+(m-1)2^q+(2m-1)2^{n-2}
.$$

Clearly, in these situations, close estimations and approximations of state complexities are usually good enough to use.

\subsubsection{Estimation of State Complexity of Combined Operations}

An estimation method through nondeterministic state complexity to
obtain the upper bound was first introduced by Salomaa and Yu~\cite{SaYu07}.
 Assume we are considering the combination of a language operation
$g_1$ with $k$ arguments together with operations  $g_2^i$, $i = 1,
\ldots, k$. The {\em nondeterministic estimation upper bound,} or
{\em NEU-bound\/} for the deterministic state complexity of the
combined operation $g_1(g_2^1, \ldots, g_2^k)$ is calculated as
follows:

\begin{itemize}
\item[{\rm (i)}]
Let the arguments of the operation $g^i_2$ be \dfas $A^i_j$ with
$m^i_j$ states,  $i = 1, \ldots, k$, $j = 1, \ldots, r_i$, $r_i \geq
1$.
\item[{\rm (ii)}] The nondeterministic
state complexity of the combined operation is at most the
composition of the individual state complexities, and hence the
language
$$
g_1( g_2^1(L(A^1_1), \ldots, L(A^1_{r_1})), \ldots, g_2^k(L(A^k_1),
\ldots, L(A^k_{r_k})))
$$
has an \nfa with at most
$$
{\rm nsc}(g_1)( {\rm nsc}(g^1_2)(m^1_1, \ldots, m^1_{r_1}), \ldots,
{\rm nsc}(g^k_2)(m^k_1, \ldots, m^k_{r_k}))
$$
states, where {\rm nsc}$(g)$ is the nondeterministic state
complexity (as a function) of the language operation $g$.
\item[{\rm (iii)}] Consequently,
the deterministic state complexity of the combined operation
$g_1(g_2^1, \ldots, g_2^k)$ is upper bounded by
\begin{equation}
\label{NEU} 2^{ {\rm nsc}(g_1)( {\rm nsc}(g^1_2)(m^1_1, \ldots,
m^1_{r_1}), \ldots, {\rm nsc}(g^k_2)(m^k_1, \ldots, m^k_{r_k})) }
\end{equation}
\end{itemize}

Table~\ref{tab:sc-neu} shows the state complexities and their
corresponding NEU-bounds of the four combined operations~\cite{SaYu07}: (1) star of union, (2) star of intersection, (3)
star of catenation, and (4) star of reversal.
\begin{table}[htbp]
\begin{tabular}{|l||c|c|}
\hline
\multicolumn{3}{|c|}{Regular}\\
\hline
&\multicolumn{1}{c}{sc}&\multicolumn{1}{c|}{NEU-bound}\\
\hline
\hline
$(L_1\cup L_2)^*$ & $2^{m+n-1}-2^{m-1}-2^{n-1}+1$
& $2^{m+n+2}$\\
\hline $(L_1\cap L_2)^*$ & $ 3/4\; 2^{mn}$
& $2^{mn+1}$ \\
\hline $(L_1L_2)^*$ & $2^{m+n-1}+2^{m+n-4}-2^{m-1}-2^{n-1}+m+1$ &
$2^{m+n+1}$ \\
\hline
$(L_1^R)^*$ & $2^m$ & $2^{m+2}$\\
\hline
\end{tabular}
\centering
  \caption{\small{State complexities of four combined operations and their
corresponding NEU-bounds on regular languages \cite{GaYu12}}}\label{tab:sc-neu}
\end{table}
This method works well when a
combined operation ends with the star operation. However, it
does not work well in general for combined operations that are ended
with reversal \cite{EsGaLiYu09,SaYu07}. For example, the state complexity of $(L(A)\cap L(B))^*$ is $2^{m + n} - 2^{m} - 2^{n} + 2$, where $A$ and $B$ are $m$-state and $n$-state \dfas, respectively. But using the above method, we would
obtain an estimate $2^{mn + 1}$. We note that in this particular case if reversal is distributed over intersection we can again recover a good estimate. Thus, it may be possible to have a general estimation method that takes in account algebraic properties of the considered model.\footnote{This observation was made to us by an annonymous referee.}

\subsubsection{Approximation of State Complexity of Combined Operations}

Although an estimation of the state complexity of a combined
operation is simpler and more convenient to use, it does not show
how close it is to the state complexity. To solve this
problem, the concept of approximation of state complexity was
proposed by Gao and Yu~\cite{GaYu09}.
The idea of approximation of state complexity comes from the notion of
approximation algorithms \cite{GaGrUl72,Jo72,Jo73}. A large number of
polynomial-time approximation algorithms have been proposed for many NP-complete problems, e.g. the
traveling-salesman problem, the set-covering problem, and
the subset-sum problem, etc. Since it is considered intractable to obtain an optimal solution for an
NP-complete problem, near optimal
solutions obtained by approximation algorithms are often good enough to use in practice. Assume there is a maximization or a minimization problem. An
approximation algorithm is said to have a ratio bound of $\rho(n)$
if for any input of size $n$, the cost $C$ of the solution produced
by the algorithm is within a factor of $\rho(n)$ of the cost $C^*$
of an optimal solution \cite{CoLeRi90:_Intro_to_algorithms}:
$$\max\left(\frac{C}{C^*}, \frac{C^*}{C}\right) \leq \rho(n).$$
The concept of approximation of state complexity is
similar to that of approximation algorithms. An approximation of state complexity of an operation
is a close estimation of the state complexity of the operation with a ratio bound showing the error range of the
approximation \cite{GaYu09}. In spite of similarities, there are some fundamental differences
between an approximation
algorithm and approximation of state complexity. The efforts in the area of approximation algorithms are
in designing polynomial algorithms for NP-complete problems such that
the results of the algorithms approximate the optimal results whereas the efforts in approximation of state complexity are in
searching directly for the estimations of state complexities such
that they are within some certain ratio bounds \cite{GaYu09}. The aim of designing an
approximation algorithm is to transform an intractable problem into
one that is easier to compute and the result is not optimal but still acceptable. In comparison, an approximation of state complexity may
have two different effects: 
\begin{enumerate}[(1)]
\item it gives a reasonable estimation of a
certain state complexity, with some bound, the exact value of which
is difficult or impossible to compute; or 
\item it gives a simpler and
more comprehensible formula that approximates a known state
complexity \cite{GaYu12}.
\end{enumerate}
Gao \emph{et al.} gave a formal definition of approximation of state complexity in \cite{GaYu12}. Let $\xi$ be a combined operation on $k$ regular languages. Assume that the state complexity of $\xi$ is $\theta$. We say that $\alpha$
is an approximation of the state complexity of the operation $\xi$
with the ratio bound $\rho$ if, for any large enough positive
integers $n_1, \ldots, n_k$, which are the numbers of states of the
\dfas that accept the argument languages of the operation,
respectively,
$$\max\left(\frac{\alpha(n_1, \ldots, n_k)}{\theta(n_1, \ldots,
n_k)}, \frac{\theta(n_1, \ldots, n_k)}{\alpha(n_1, \ldots,
n_k)}\right) \leq \rho(n_1, \ldots, n_k).$$ Note that in many cases,
$\rho$ is a constant.
 Some examples of approximation of state complexity of combined operations are shown in Table~\ref{tab:approximation}.

\begin{table}[htbp]
\begin{tabular}{|l||c|c|}
\hline
\multicolumn{3}{|c|}{Regular}\\
\hline
&\multicolumn{1}{c}{Approximation}&\multicolumn{1}{c|}{Ratio bound}\\
\hline
\hline
$(L_1\cup L_2)^*$ & $2^{m+n+2}$  & $\approx 8$  \cite{GaYu12}\\
\hline $(L_1\cap L_2)^*$ & $2^{mn+1}$ & $8/3$  \cite{GaYu12}\\
\hline $(L_1L_2)^*$  &$2^{m+n+1}$ & $\approx 4$   \cite{GaYu12}\\
\hline
$(L_1^R)^*$ & $2^{m+2}$ & $4$  \cite{GaYu12}\\
\hline
$(L_1\backslash R)^*$ & $2^{m-1}+2^{m-2}$ & $\frac{4}{3}$ \cite{GaYu09} \\
\hline
$L_1\backslash R^*$ & $2^{m+1}$ & $\frac{8}{3}$ \cite{GaYu09} \\
\hline
\end{tabular}
\centering
  \caption{\small{Approximations of state complexities of six combined operations and their
corresponding ratio bounds on regular languages}}\label{tab:approximation}
\end{table}


\section{Conclusions}
\label{sec:conclusions}
In the last two decades, a huge amount of results were obtained on operational state complexity of regular languages.
Results are roughly  split between: individual and combined operations; regular and different classes of subregular languages; deterministic and nondeterministic complexity; different alphabet sizes; and worst case versus average case. 
In general, all this work also suggest new directions of research and open problems.
 
As it is evident by this survey, 
many results on this area are functions parametrized by some measures, mostly the state complexities of the operation arguments. Given the amount and diversity of these functions, it is useful to have a software tool that helps to structurally organize, visualize and manipulate this information. Towards this goal,  a first step was taken by the development of \textsf{DesCo},  a Web-based information system  for descriptional complexity results~\cite{desco,nabais13a:_desco}. \textsf{DesCo} keeps information about language classes, languages operations, models of computation, measures of complexity and complexity functions (both operational and transformational). 
For instance, given an operation, it is possible to obtain 
the complexity functions for all language classes and all complexity measures (that are registered in the database).

To obtain a witness for a tight upper bound, many authors performed experiments using computer software. The reason why some witnesses would work for several  (or almost all) complexity bounds only recently has been addressed.
Universal witnesses (and their variants) for operational state complexity of regular languages can be considered a major breakthrough. Conditions for a family of languages  to be \emph{universal} include also other measures as the syntactic complexity and the number of atoms.
The study of necessary and/or sufficient conditions for the maximality of all these measures is a new direction of research. Other open problems are how and whether this approach  extends to other classes of subregular languages and to other complexity measures, in particular to  nondeterministic state complexity and transition complexity.

Besides the worst-case complexity  of an operation, researchers also studied the range of possible values that can be achieved, as a function of the complexities of the arguments and the alphabet size. A magic value  is a value that cannot occur (for that kind of complexity, operation and alphabet size). In general, if growing alphabet sizes are allowed no magic numbers exists (and even for binary alphabets they are rare). The distribution of possible complexity values and the density of languages (or tuples of languages) that achieve that values can also be valuable for average-case analysis. 

Witnesses with alphabets of increasing size were used in the quest of magic numbers, for the state complexity of certain operations over subregular languages, and almost for all results on combined operations with an arbitrary number of operands.
This suggest the question of whether the alphabet size should be a parameter of the  complexity under study. In particular, it should be investigated which situations cannot be characterized without increasing alphabets, and the ones for which languages with fixed alphabets can exists but are not  yet known.

For many automata applications, a major direction of research is average-case state complexity. An essential question for average results is the probability distribution  that is chosen for the models. The few results that exist use a uniform distribution, and even in this case the problem is very difficult. Recently, using the framework of analytic combinatorics, some average-case results were obtained for the size of \nfas equivalent to a given regular expression~\cite{nicaud09a:_averag_size_of_glush_autom,broda11:_averag_state_compl_of_partial_deriv_autom,broda12:_averag_size_of_glush_and,broda13:_hitch_guide_descr_compl_analy_combin}. It is also worthwhile to mention the average-case computational complexity analysis of the Brzozowski minimization algorithm carried on by Felice and Nicaud~\cite{felice13:_brzoz_algor_is_gener_super,felice14:_averag_compl_of_brzoz_algor}. This work can be specially relevant for the operational state complexity because the authors give some  characterizations of the state complexity of reversal. Another approach for average-case analysis is to consider  experimental results based on samples of uniformly random generated automata. There are some random generators for non-isomorphic 
 \dfas\cite{almeida07_c:_enumer_gener_strin_autom_repres,bassino07:_theor_comput_scien,felice13:_random_gener_of_deter_acycl}, but for \nfas, the fact that  there is no generic polynomial algorithm for graph isomorphism, the problem seems unfeasible in general.

\section{Acknowledgements}
\label{sec:ack}
This work was partially supported by the
    European Regional Development Fund through the programme COMPETE
    and by the Portuguese Government through the FCT under 
    projects PEst-C/MAT/UI0144/ 2013 and PTDC/EIA-CCO/101904/2008.
The writing of this paper started in the fall of 2010 when Nelma Moreira
and Rogério Reis were visiting Sheng Yu at University of Western
Ontario. When Sheng passed away in January 2012, the paper was already
almost in the current form.  We deeply thank Kai Salomaa for his
encouragement, suggestions, and corrections. We also thank Davide
Nabais and Eva Maia for proofreading.
Comments and criticisms of the anonymous referee were most valuable for improving the 
final version of the paper.


%
\section{References}
\bibliography{sc}{}
\bibliographystyle{plain}
\end{document}